\documentclass[12pt,letterpaper,a4paper]{article}
\usepackage[includeheadfoot,
            marginratio={1:1,2:3},
            width=412pt,
            height=688pt,]{geometry}
\usepackage{amsmath}
\usepackage{amsfonts}
\usepackage{amssymb}
\usepackage{graphicx}
\usepackage{empheq}
\usepackage{color}
\usepackage{hyperref}

\newcommand{\nc}{\newcommand}
\nc{\beq}{\begin{equation}}
\nc{\eeq}{\end{equation}}
\nc{\bea}{\begin{eqnarray}}
\nc{\eea}{\end{eqnarray}}
\def\ov{\overline}
\def\cO{{\cal O}}
\def\IP{\mathbb{P}}

\newdimen\csize\csize=1.5ex
\def\young#1{\tiny\vcenter{\hbox{\vrule\vtop{\hrule
  \offinterlineskip\halign{&\vbox
  {\hbox to\csize {\strut\hss##\hss\vrule}\hrule}\cr#1 \crcr}}}}}

\newcommand{\eq}[1]{\begin{equation}
                     \begin{split} #1 \end{split}
                     \end{equation}}

\begin{document}

\vspace*{-1.5cm}
\begin{flushright}
  {\small
  MPP-2012-123\\
  }
\end{flushright}

\vspace{1.5cm}
\begin{center}
  {\Large
Moduli Stabilization and Inflationary Cosmology  with \\[0.2cm]
Poly-Instantons in Type IIB Orientifolds }

\end{center}

\vspace{0.75cm}
\begin{center}
  Ralph Blumenhagen$^1$, Xin Gao$^{1,2}$, Thorsten Rahn$^1$ and
Pramod Shukla$^1$
\end{center}

\vspace{0.1cm}
\begin{center}
\emph{
$^{1}$ Max-Planck-Institut f\"ur Physik (Werner-Heisenberg-Institut), \\
   F\"ohringer Ring 6,  80805 M\"unchen, Germany\\
\vskip 1cm
$^{2}$ State Key Laboratory of Theoretical Physics, \\
Institute of Theoretical Physics,\\ Chinese Academy of Sciences, P.O.Box 2735, Beijing 100190, China } \\[0.1cm]

\vspace{0.2cm}

 \vspace{0.5cm}
\end{center}

\vspace{1cm}


\begin{abstract}
Equipped with concrete examples of
Type IIB orientifolds featuring poly-instanton corrections
to the superpotential, the effects
on moduli stabilization and inflationary cosmology are analyzed.
Working in the framework of the  LARGE volume scenario, the K\"ahler
modulus related to the size of the four-cycle supporting
the poly-instanton contributes sub-dominantly to the scalar potential.
It is shown that this K\"ahler modulus  gets stabilized and,
by displacing it from its minimum, can play the role of an inflaton.
Subsequent cosmological implications are discussed
and compared to  experimental data.
\end{abstract}

\clearpage



\section{Introduction}
\label{sec:intro}
For relating string compactifications to realistic four-dimensional physics,
the understanding of moduli stabilization is a  very central issue and has
been under intense investigation  for more than a decade. In order to
quantitatively address certain  aspects of cosmology and  of particle physics,
moduli stabilization is a prerequisite, as on the one hand some physical
parameters depend on the value of the moduli and on the other hand
the existence of unstabilized and therefore  massless scalars is incompatible
with observations.

Motivated by these issues, quite some progress concerning mechanisms to
stabilize these moduli has been made over the past years
\cite{Grana:2005jc,Blumenhagen:2006ci,Gukov:1999ya,Kachru:2003aw,Balasubramanian:2005zx,Bobkov:2010rf}. In
the framework of Type IIB orientifold compactifications with $O7$ and
$O3$-planes, one can distinguish
two classes of models, namely the so-called  KKLT models \cite{Kachru:2003aw}
and the LARGE volume scenarios \cite{Balasubramanian:2005zx}. Both of them
successfully stabilize the moduli via (non-)perturbative effects through a
two-step strategy. First, one stabilizes the complex structure moduli as
well as the axio-dilaton modulus via a dominant perturbative
contribution.
Second, the K\"ahler moduli are stabilized via a sub-dominant
non-perturbative contribution.

More concretely, in KKLT models, the complex structure moduli as well as the
axio-dilaton are stabilized via contributions to the
superpotential that are generated by turning on tree-level background
fluxes \cite{Gukov:1999ya}. On the other hand the
K\"{a}hler moduli are stabilized via non-perturbative effects coming from
gaugino-condensation or Euclidean $D3$-brane (short $E3$-brane)
instantons \cite{Witten:1996bn}. These effects
are sufficient to fix all closed string moduli (at least for simple setups)
in  a supersymmetric $AdS$ minimum, which can be uplifted to a
non-supersymmetric metastable $dS$ minimum by  the inclusion of
$\overline{D3}$-branes \cite{Kachru:2003aw}.

Though being very nice and simple, a problematic  issue is that KKLT models
suffer from a  lack of control over additional corrections.
In particular, this is improved in the LARGE volume
scenarios \cite{Balasubramanian:2005zx},
where the inverse of the volume serves as a controllable expansion
parameter.
Here, one includes
perturbative ${\alpha^\prime}^3$-corrections (of \cite{Becker:2002nn}) in the
K\"{a}hler potential, leading to a new non-supersymmetric
$AdS$ minimum
where the modulus controlling the size of the instanton $\tau_s$ and
the overall volume modulus ${\cal V}$
are stabilized at hierarchically separated values such that
$\tau_s\sim \ln{\cal V}$.
Thus, the  basic idea   is to balance a
non-perturbative correction to the superpotential against a
perturbative correction to  the K\"{a}hler potential.
Unlike KKLT models, even for generic choices of background fluxed with
$W_0 \sim {\cal O}(1)$, the overall Calabi-Yau volume is sufficient to
suppress the higher derivative effects and at the same time to stabilize
(all) divisor volumes.
The $AdS$-minima in the large volume limit can  then be uplifted
making use of any of the known  uplifting mechanisms \cite{Kachru:2003aw,Westphal:2006tn,Burgess:2003ic,Saltman:2004sn,Cicoli:2012fh}.

Equipped with such  a working framework of moduli stabilization, one is conceptually able
to also address cosmological issues like realizing inflation.
The central requirement for this purpose is to identify a scalar which could
play the role of the inflaton field, i.e.\ its effective scalar potential has
to admit a slow-roll region. In this respect, those ``moduli'' that have a
flat potential at leading order and only by a sub-leading effect receive
their dominant contribution are of interest.
Besides sub-leading perturbative effects also instanton effects can be
relevant(see \cite{Gukov:1999ya,Witten:1996bn,Becker:2002nn,Blumenhagen:2009qh,Grimm:2011dj} and references therein).

Although  the concept of inflation has been proposed quite some time ago as a
solution to certain cosmological problems \cite{Guth:1980zm,Linde:1981mu}, the
embedding of inflationary scenarios into  semi-realistic string models is
quite a recent development. Such an inflationary model has been initiated in \cite{Kachru:2003sx} in which, following the idea of \cite{Dvali:1998pa}, an open string modulus appearing as brane separation was argued to be an inflaton candidate. There has been a large amount of work dedicated to build sophisticated models of open string inflation \cite{Kachru:2003sx,Dasgupta:2004dw,Baumann:2009qx}, however in particular in the framework of LARGE volume scenarios, closed string moduli inflation has also been seriously considered with a successful and consistent implementation of the observational data \cite{Conlon:2005jm,Conlon:2008cj,Cicoli:2008gp,Cicoli:2011zz}.
Models based on the $C_4$ axion-moduli have also been considered for
cosmological implications
\cite{Dimopoulos:2005ac,BlancoPillado:2004ns,BlancoPillado:2006he}, whereas
models with odd-axions $B_2, C_2$, though also  interesting, are  less studied
\cite{Kallosh:2007cc,Grimm:2007hs,Misra:2007cq,McAllister:2008hb}. All these
string (inspired) inflationary models experience significant constraints from
experimental observations, i.e.\ from the measurements  of the temperature
fluctuation in the cosmic microwave background (CMB). This includes
the COBE,  WMAP and Planck-satellite missions.

Along the lines of moduli getting lifted by  sub-dominant contributions,
recently so-called poly-instanton corrections became  of interest.
These are  sub-leading non-perturbative contributions which can be briefly
described as instanton corrections to instanton actions. These were introduced
and systematically studied in the framework of Type I string compactifications
in  \cite{Blumenhagen:2008ji} and have also been analyzed  in the 
T-dual context of Type I' string models, where they map to 
BPS  $D0$-branes running in loops \cite{Petersson:2010qu}. 
The analogous poly-instanton effects will also
appear in the Type IIB orientifolds with $O7$ and $O3$ planes.
Utilizing these  poly-instanton effects, moduli stabilization and inflation
have been studied in a series of papers
\cite{Blumenhagen:2008kq,Cicoli:2011yy,Cicoli:2011ct,Cicoli:2012cy,Cicoli:2012tz}.
However the analysis had to be  carried out from a rather heuristic point of
view, as a clear understanding of the string-theoretic conditions  for the
generation of these effects was lacking. In this context,
in the recent work
\cite{Blumenhagen:2012Poly1}, we have clarified
the zero mode conditions for an Euclidean D3-brane instanton, wrapping a
divisor of the threefold, to  generate such a poly-instanton effect.
There we have also  provided the construction of some concrete
models having the right divisors to both feature the
LARGE volume scenario and to support additional poly-instanton corrections.
Building on these,
it is the objective of this paper to study its consequences for
moduli stabilization and inflation.

The article is organized as follows. In section \ref{sec_Poly-instanton
  corrections}, we start with a brief review of the relevant mathematical
background (from \cite{Blumenhagen:2012Poly1}) exemplifying it
for a concrete Type IIB orientifold model in which poly-instanton corrections are generated. In
section \ref{sec_Moduli stabilization} we present a systematic
study of moduli stabilization. Here we distinguish two schemes, a so-called
minimal one and one with a racetrack form of the superpotential.
A numerical analysis leads to the existence of critical points and
a hierarchy of moduli masses.
In section \ref{sec_Inflationary Cosmology} we analyze  the potential
of the  lightest modulus to  serve as an inflaton. For this purpose,
we compute several inflationary parameters and compare them to the
observed values. Moreover, we give a first, rough  estimate of the
reheating temperature.
Finally in section \ref{sec_Conclusions and Discussions} we give our
conclusions followed by an appendix providing   some more details
on the partially quite lengthy  intermediate expressions.

\section{Poly-instanton corrections}
\label{sec_Poly-instanton corrections}

Let us briefly recall some results from \cite{Blumenhagen:2012Poly1} on the
contribution  of poly-instantons to the superpotential in the
framework of Type IIB  orientifold
compactifications on Calabi-Yau threefolds with $O7$- and $O3$-planes\footnote{For a general review on D-brane instantons effects see
\cite{Blumenhagen:2009qh}.}.
In this case  the orientifold action is given by $\Omega \sigma (-1)^{F_L}$,
where $\sigma$ is a holomorphic, isometric involution acting
on the Calabi-Yau threefold ${\cal M}$.

Generally, the notion of poly-instantons \cite{Blumenhagen:2008kq} means the correction
of an Euclidean D-brane instanton action  by other D-brane instantons.
The configuration of interest in the following is that we have two
instantons $a$ and $b$ with
proper zero modes to generate a  non-perturbative contribution
to the superpotential of the form
\eq{
W=A_a\, \text{exp}^{-S_a}+A_a A_b\, \text{exp}^{-S_a-S_b}+ ...\, ,
}
where $A_{a,b}$ are moduli dependent one-loop determinants and $S_{a,b}$
denote the classical $D$-brane instanton actions.
The latter  depend on different moduli due to the different zero mode structures
for these two kinds of instantons.

In the framework of Type IIB  orientifolds,
sufficient conditions for the  zero mode structures
for  poly-instanton corrections to the superpotential  have been
worked out  in \cite{Blumenhagen:2012Poly1}.
Both instantons, $a$ and $b$, should be $O(1)$ instantons,
i.e.\ a single instanton placed in an orientifold invariant position with
an $O(1)$ projection.  This  corresponds to an $SP-$type projection
for a corresponding  space-time filling $D7$-brane.
Instanton $a$ is an Euclidean $E3$ instanton wrapping a rigid divisor $E$ in
the Calabi-Yau threefold  with $H^{1,0}(E,{\cal O})=H^{2,0}(E,{\cal O})=0$.
Whereas, instanton $b$ is  an Euclidean $E3$-brane  instanton
wrapping a divisor which admits a single complex Wilson line Goldstino,
i.e.\ a so-called  Wilson line divisor with equivariant
cohomology $H^{*,0}(W,{\cal O})=(1_+,1_+,0)$ under the involution $\sigma$.
Here the lower indices $\pm$ denote the even and odd cohomology.
In fact, the sufficient condition for a geometric configuration to support the
poly-instanton correction is that it precisely contains one
Wilson line modulino in $H^1_+(E,\cal O)$.

Examples of such Wilson line divisors are $\IP^1$ fibrations
over two-tori. In \cite{Blumenhagen:2012Poly1} a couple
of concrete Calabi-Yau threefolds,
both with and without $K3$ fibration structure, were presented which
featured all the requirements mentioned above.
For concreteness, let us recall one of these examples, which
not only admits a Wilson line divisor $W$ but also two rigid and
shrinkable del Pezzo divisors. Such geometries are particularly
interesting for studying moduli stabilization as they give rise to a
swiss-cheese type K\"ahler potential, the starting point
for the LARGE volume scenario \cite{Balasubramanian:2005zx}.
The Calabi-Yau threefold $\cal M$ is given by a hypersurface
in a toric variety with defining data
\begin{table}[ht]
  \centering
  \begin{tabular}{c|cccccccc}
     & $x_1$  & $x_2$  & $x_3$  & $x_4$  & $x_5$ & $x_6$  & $x_7$ & $x_8$        \\
    \hline
    2  & -1 & 0 & 1 & 1 & 0 & 0 &  0 & 1  \\
    4  & -2 & 0 & 2 & 2 & 1 & 0 &  1 & 0  \\
    2  & -3 & 0 & 2 & 1 & 1 & 1 &  0 & 0  \\
    2  & 1 & 1 & 0 & 0 & 0 & 0 &  0 & 0  \\
  \end{tabular}
 \end{table}

\noindent
with Hodge numbers $(h^{21}, h^{11}) = (72, 4)$ and the corresponding Stanley-Reisner ideal
\begin{equation}
{\rm SR}=  \left\{x_1\,x_2,  x_4\,x_7,  x_5\,x_7,  x_1\,x_4\,x_8,
x_2\,x_5\,x_6,  x_3\,x_4\,x_8,  x_3\,x_5\,x_6,  x_3\,x_6\,x_8 \right\}\,.
\end{equation}
More geometric data can be found in \cite{Blumenhagen:2012Poly1}.

We identified  two inequivalent orientifold projections $\sigma: \{x_4
\leftrightarrow -x_4, x_7 \leftrightarrow -x_7\}$ with $h^{11}_-({\cal M})=0$
so that for the Wilson line divisor $W=D_8=\{x_8=0\}$ 
the Wilson line Goldstino is in $H_{+}^1(W,\cO)$. It was
checked that the $D3$- and $D7$-brane tadpoles can be canceled. For the
analysis in the following sections, we focus on the
involution $x_7 \leftrightarrow -x_7$.
The corresponding topological data of the relevant divisors are shown
in table \ref{tabledivsA}.
\begin{table}[ht]
  \centering
  \begin{tabular}{c|c|c}
    divisor & $(h^{00},h^{10},h^{20},h^{11})$  &  intersection curve    \\
    \hline \hline
      &    &  \\[-0.4cm]
    $D_7=dP_7$ & $(1_+,0,0,8_+)$ & $W: C_{g=1}$ \\
    $D_5$ & $(1_+,0,1_+,21_+)$ & $W: C_{g=1}$ \\
    \hline
    $D_8=W$ & $(1_+,1_+,0,2_+)$ & $D_5: C_{g=1}, \ \ D_7:  C_{g=1}$ \\
    $D_1=\IP^2$ & $(1_+,0,0,1_+)$ & $D_5: C_{g=0}$
  \end{tabular}
  \caption{\small Divisors and their equivariant cohomology under $x_7 \leftrightarrow -x_7$. The first two
    lines are $O7$-plane components  and the remaining two divisors
    can support $E3$ instantons. The $D_7$ divisor also supports
    gaugino condensation.}
  \label{tabledivsA}
\end{table}

\noindent
We also showed  that
 there are no extra vector-like zero modes on the intersection of $E3\cap D7$,
i.e.\  all sufficient conditions were satisfied for the divisor $W$
to generate a poly-instanton correction to the  non-perturbative superpotential
\eq{
   W&= A_1\, \exp\left( -2\pi T_1\right) + A_1\, A_8\, \exp\left(
       -2\pi T_1-2\pi T_8\right)+\\
&\phantom{aaaaaaaaaaa}A_7\, \exp\left( -a_7 T_7\right) + A_7\, A_8\, \exp\left(
       -a_7 T_7-2\pi T_8\right)+\ldots\; .
}

\noindent
Taking into account the K\"ahler cone constraints, the volume form for this
model could  be written in the strong swiss-cheese like form
\eq{
\label{volumeA}
{\cal V}&=\textstyle{ \frac{1}{9}}\Bigl(\frac{1}{\sqrt 2}
(\tau_1+3\tau_6+6\tau_7+3\tau_8)^{3/2}-\sqrt{2}\tau_1^{3/2}-3\tau_7^{3/2}-3(\tau_7+\tau_8)^{3/2}\Bigr)\,.
}
The above volume form shows that the large volume limit is given by
 $\tau_6 \rightarrow \infty$ while keeping the other shrinkable del Pezzo
four-cycles $\tau_{1,7}$ and the Wilson line four-cycle $\tau_8$ small.
 Note, all other models studied in \cite{Blumenhagen:2012Poly1} shared a
similar strong swiss-cheese-like volume form with the same intriguing
appearance of the Wilson line K\"ahler modulus.

These concrete examples motivate us now to make the following
slightly simplified ansatz for the tree-level K\"ahler potential
and the poly-instanton  generated superpotential
\eq{
\label{kaehlerex}
   K=-2\ln {\cal V}= -2 \ln\left( {\tau_b}^{\frac 3 2}
                 - {\tau_s}^{\frac 3 2}
                -(\tau_s+\tau_w)^{\frac 3 2}\right)\,,}
\eq{
\label{superex}
     W =A_s\, e^{-a_s T_s} + A_s \, A_w\, \, e^{-a_s \,T_s- a_w\,T_w}\,,
}
where $A_{s,w}$ are one-loop determinants and $a_{s,w}$ are
numerical constants.

It is the objective of this paper to analyze the physical implications
of such a K\"ahler  and superpotential
for moduli stabilization  and inflationary cosmology.

\section{Moduli stabilization}
\label{sec_Moduli stabilization}

A generic orientifold compactification of Type IIB string theory
leads to an effective four-dimensional ${\cal N}=1$ supergravity theory.
In the closed string sector, the bosonic part of the 
massless chiral superfields arises from
the dilaton, the complex structure and  K\"ahler moduli and
the dimensional reduction of the NS-NS and R-R $p$-form fields.
The bosonic field content is given by
\eq{
\label{eq:N=1_coords}
 \tau&=C^{(0)}+ie^{-\phi} \, , \qquad  U^i = u^i + i v^i, \quad
   i=1\ldots h^{21}_+ \,, \\[0.1cm]
 {G}^a & = c^a - \tau {b}^a \,, \qquad a=1,\ldots, h^{11}_- \,, \\
 T_\alpha&=\frac{1}{2} \kappa_{\alpha\beta\gamma}t^\beta t^\gamma +
 i\left(\rho_\alpha -\kappa_{\alpha a b} \, {c^a b^b}\right) +\frac{i}{2}\,
 \tau \kappa_{\alpha ab} b^a b^b \quad \text{~and} \quad  \alpha=1,\ldots, h^{11}_+\,,
}
where $c^a$ and $b^a$ are defined as integrals of the
axionic $C^{(2)}$ and $B^{(2)}$ forms and $\rho_{\alpha}$ as
integrals of $C^{(4)}$ over a basis of four-cycles $D_\alpha$.
From now on, as in our concrete examples, we assume $h^{11}_-=0$.

The supergravity action is specified by the K\"ahler potential,
the holomorphic superpotential $W$ and the holomorphic gauge kinetic function.
The K\"{a}hler potential for the supergravity action is given as,
\begin{eqnarray}
\label{eq:K}
& & \hskip -1cm K = - \ln\biggl(-i(\tau-{\bar\tau})\biggr)
-\ln\left(-i\int_{\cal M}\Omega\wedge{\bar\Omega}\right)-2\ln\Bigl( {\cal V}(T_\alpha)\Bigr)\,,
\end{eqnarray}
where ${\cal V}={\frac 1 6} {\kappa}_{\alpha\beta\gamma} t^\alpha t^\beta t^\gamma$
is   the volume of the internal
Calabi-Yau threefold.

The general form of the superpotential $W$ is given as
\begin{equation}
\label{eq:W}
W = \int_{\cal M} G_3\wedge\Omega + \sum_{E} {A}_{E}(\tau, U^i) \,
e^{- a_E \gamma^\alpha\, T_{\alpha}} \,
\end{equation}
with the instantonic divisor given by $E=\sum  \gamma^\alpha D_\alpha$.
The first term is the Gukov-Vafa-Witten (GVW)  flux induced
superpotential \cite{Gukov:1999ya}  and the second one denotes  
the non-perturbative
correction coming from Euclidean $D3$-brane instantons ($a_E=2\pi$) and
gaugino condensation on $U(N)$ stacks of $D7$-branes ($a_E=2\pi/N$).
In terms of the  K\"{a}hler potential and the superpotential
the scalar potential is given by
\eq{
\label{eq:Vgen}
V = e^{K}\Biggl(\sum_{I,\,J}{K}^{I\bar{J}} {\cal D}_I W {\bar{\cal D}}_{\bar J } {\bar W} - 3 |W|^2 \Biggr)\,,
}
where the sum runs over all moduli.
As usual in the  LARGE volume scenario, the complex structure moduli
and the axio-dilaton
are stabilized  at order $1/{\cal V}^2$ by the GVW-superpotential.
Since the stabilization of the K\"{a}hler moduli is by sub-leading
terms in the ${\cal V}^{-1}$ expansion, for this purpose
the complex structure moduli and the dilaton can be treated as
constants.

In the remainder of this section we will analyze
moduli stabilization for an effective supergravity model
defined essentially  by the tree-level K\"ahler
potential \eqref{kaehlerex} along with perturbative ${\alpha^\prime}^3$ correction and the non-perturbative
superpotential \eqref{superex}.
We will discuss two schemes for  moduli stabilization.
In the first one, up to an additional flux induced constant
$W_0$,  we will  just consider
the minimal superpotential \eqref{superex} and in the second one we will
extend the model  to a racetrack-type variant.

We will see that the first scheme is problematic in the sense
that the minima of the scalar potential lie outside the
regime of validity of the $\alpha'$ expansion.
This is remedied in the second scheme, with the downside that one has to
perform a certain tuning of the parameters.

\subsubsection*{Scheme 1: minimal}

Note that, for $h^{11}_{-} =0$, the ${\cal N} = 1$ K\"{a}hler
coordinates are simply given as $T_\alpha = \tau_\alpha + i \rho_\alpha$.
For the K\"{a}hler and superpotential, we choose  the
minimal poly-instanton motivated  form
\eq{
\label{eq:KW+no_race}
 K &= - 2 \, \ln {\cal Y} \,, \\[0.2cm]
 W &= W_0 + A_s e^{-a_s T_s}+ A_s A_w\, e^{-a_s T_s - a_w T_w}\,,
}
where ${\cal Y}= {\cal V}(T_\alpha)+C_{\alpha^\prime}\;$ such that
\eq{
{\cal Y}
= \xi_b (T_b+\bar{T}_b)^{\frac{3}{2}}-\xi_s
(T_s+\bar{T}_s)^{\frac{3}{2}}-\xi_{sw}
\Bigl((T_s+\bar{T}_s) + (T_w+\bar{T}_w)\Bigr)^{\frac{3}{2}} +
C_{\alpha^\prime}\, .
}
Here, $C_{\alpha^\prime}$ denotes  the perturbative
${\alpha^\prime}^3$-correction  given as
\eq{
C_{\alpha^\prime} = - \frac{\chi({\cal M}) \,
  ({{\tau}-\bar\tau})^{\frac{3}{2}} \zeta(3) }{4 (2 \pi)^3 \,
  ({2i})^{\frac{3}{2}}}
}
with $\chi({\cal M})$ being the Euler characteristic of the Calabi-Yau.
The large volume limit is defined by taking $\tau_b \rightarrow \infty$ while
keeping the other divisor
volumes  small. Note that the above volume form is very similar to
the one of  `strong' swiss-cheese type.

\begin{figure}[!ht]
\centering

\includegraphics[scale=0.3]{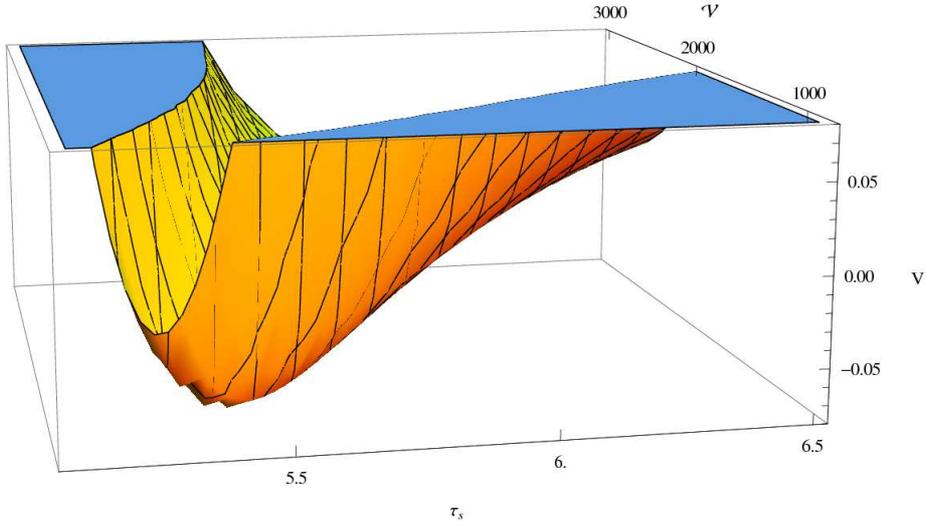}
\caption{The scalar potential $V({\cal V},\tau_s)$ (multiplied by $10^{7}$)
  as a function of  ${\cal V}$ and $\tau_s$  in the absence of poly-instanton effects. The model dependent parameters have been chosen to be $\chi({\cal M}) = -136$, $\xi_b = \frac{1}{36}$, $\xi_s  = \xi_{sw} =\frac{1}{6\sqrt2}$, $a_s = \frac{2\pi}{4}$, $A_s =8$, $g_s= 0.12$, $W_0 = -5$. The respective minimum values for the moduli are $\ov\rho_s=0$, $\ov\tau_s = 5.44$, $\ov{\cal V} = 1086$. 
 }\label{S11}
\end{figure}

The effective scalar potential for the set of moduli
$\{T_b, T_s, T_w\}$ can be expressed  in terms of  three
types of contributions
\eq{
V(\tau_b,\tau_s,\tau_w,\rho_s,\rho_w) \equiv V({\cal
  V},\tau_s,\tau_w,\rho_s,\rho_w)= V_{\alpha^\prime} + V_{\rm np1} + V_{\rm np2} \, .
}
In the large volume limit, the leading contributions are
\eq{
\label{eq:Vgen+norace}
V_{\alpha^\prime}  & = \frac{3 {\, {\cal C}_{\alpha^\prime}} \,|W_0|^2}{2
  {\cal V}^3}\, ,\\[0.2cm]
V_{\rm np1} &=\frac{4\, {W_0}}{{\cal V}^2}\biggl[{\,{A_s}} \, e^{-{a_s \tau_s}-{a_w \, \tau_w}} \Bigl({\,a_s} e^{{a_w \, \tau_w}}
{\tau_s} \cos ({\,a_s} {\rho_s})\\
&\phantom{aaaaaaaaaaaaaaaaaa}
+{\,{A_w}} ({a_s \tau_s}+{a_w \, \tau_w}) \cos ({\,a_s} {\rho_s}+{\,{a_w}}
{\rho_w})\Bigr) \biggr]\, , \\
  V_{\rm np2} &= \frac{2\sqrt2}{3 \, \xi_{s} \, \xi_{sw} \, {\cal V}} \biggl[
 A^2_s\, a_s^2\, e^{-2 ({a_s \tau_s}+{a_w \, \tau_w})}
\Bigl(\xi_{sw} \left(A^2_w+e^{2 {a_w  \tau_w}}\right) \sqrt{{\tau_s}} \\
&\phantom{aaaaaaaaaa}
+2 \xi_{sw}\, a_s\, (a_s-a_w) A_w\, e^{a_w \tau_w} \sqrt{\tau_s} \cos (a_w
  \rho_w) \\
&\phantom{aaaaaaaaaa}
 -2 \xi_{sw}\, {a_w} a_s \,A^2_w \,
\sqrt{\tau_s} +a^2_w \,A^2_w \left(\xi_s\, \sqrt{{\tau_s}+{\tau_w}}
+\xi_{sw} \sqrt{\tau_s}\right)\Bigr)\biggr]\, .
}
In the absence of poly-instanton corrections
the above scalar potential reduces to
\eq{
\label{eq:V_lvs}
V^{\rm LVS}({\cal V},\tau_s,\rho_s) =& \frac{2 {\sqrt 2}\,a_s^2 \,A_s^2\, e^{-2 {a_s \tau_s}} \sqrt{{\tau_s}}}{3 \, \xi_s \,
 {\cal V}}+\frac{{4 \,a_s} {\,{A_s}}\, e^{-{a_s \tau_s}}\, {\tau_s} \cos
  (a_s \rho_s) {W_0}}{{\cal V}^2}\\[0.1cm]
&+\frac{3 {\, {\cal C}_{\alpha^\prime}} \,|W_0|^2}{2 \, {\cal V}^{3}}
}
which corresponds to the potential of the  standard single-hole LARGE
volume scenario \cite{Balasubramanian:2005zx}.
The extrema  of the above scalar potential
can be  collectively\footnote{The  extremizing condition for the axion
can be decoupled from those of the divisor volumes.
Then, the extremality conditions $\partial_{{\cal V}} V = 0$ and
$\partial_{\tau_s} V = 0$ result in two coupled constraints
in ${\cal V}$ and $\tau_s$, which can be combined so that
in one condition (the second in \eqref{hyperextr})
$\tau_s$ and ${\cal V}$ are decoupled.}
described  by the following three hypersurfaces in moduli space
\eq{
\label{hyperextr}
a_s \ov\rho_s &= N \pi \, ,  \qquad
{\cal C}_{\alpha^\prime} = \frac{32 \sqrt{2} a_s \, \xi_s
\, \ov\tau_s^{\frac{5}{2}} (-1 + a_s \ov\tau_s)}{(-1 + 4 a_s \ov\tau_s)^2}
\,,\\[0.1cm]
W_0 &= -\frac{\, a_s A_s \,
e^{-a_s \ov\tau_s}\, \ov{\cal V} \, (-1 + 4 a_s \ov\tau_s)}{6 \, {\sqrt 2} \,
  \xi_s  \, \sqrt{\ov\tau_s}\, (-1 + a_s \ov\tau_s)}\, .
}
From the above, one finds that $\tau_s$ gets stabilized in terms
of the $\alpha^\prime$-correction term in the scalar potential, i.e.~$\ov\tau_s \sim (C_{\alpha^\prime})^{2/3}$.
Then ${\cal V}$ gets stabilized at an  exponential large
volume  $\ov{\cal V}\sim \exp(a_s\, \ov\tau_s)$.
The form of the scalar potential is shown in figure \ref{S11}.

In addition to the standard LARGE volume scenario
terms \eqref{eq:V_lvs}, the generic potential \eqref{eq:Vgen+norace}
involves sub-dominant contributions. These  are further
suppressed by powers  of $e^{-a_w \tau_w}$.
Collecting the sub-leading terms of type $e^{-a_w \tau_w}$, we find
\eq{
V(\tau_b,\tau_s,\tau_w,\rho_s,\rho_w) = V^{\rm LVS}({\cal V},\tau_s,\rho_s)+
V(\tau_w,\rho_w)\,,
}
where
\eq{
 V(\tau_w,\rho_w) \sim &  \frac{4 \, A_s \, A_w \, (a_s \tau_s + a_w \tau_w) \, e^{-a_s \tau_s -a_w \tau_w} \, W_0 \, \cos(a_s \rho_s + a_w \rho_w) }{{\cal V}^2} \\[0.1cm]
 & + \frac{2 \sqrt2 \, a_s (a_s -a_w)\, A_s^2 \, A_w\, {\sqrt\tau_s }  \,
   e^{-2 a_s \tau_s - a_w \tau_w} \, \cos(a_w \rho_w)}{3\, \xi_s \, {\cal
     V}}\, .
}
Thus, the K\"ahler  moduli $\tau_b$ and $\tau_s$ are stabilized as in
the standard LARGE volume scenario at order ${\cal V}^{-3}$.
At this order, the modulus $\tau_w$ remains a flat direction.
This makes it a natural candidate for realizing slow-roll inflation.
The Wilson-line divisor volume $\tau_w$  gets  lifted
once the sub-dominant poly-instanton effects are included.
Moreover, with lifting the flatness
via a sub-dominant correction, one usually expects that the mass-hierarchy will
be maintained during the motion of the  lightest
modulus while keeping the heavier ones at their minimum.

Now, assuming that the single field approximation is a valid description
and  after stabilizing the heavier moduli $\tau_b,\tau_s$ (and the
relevant axion moduli) at their respective minima by solving
\eqref{hyperextr},
the sub-leading scalar potential
for $\tau_w$ is given by
\bea
& &  V(\tau_w) = V^{\rm LVS}(\ov{\cal V},\ov\tau_s,\ov\rho_s) + \, e^{- a_w \tau_w}\, \bigl(\gamma_1 + \gamma_2 \tau_w\bigr)\,,
\eea
where the constants $\gamma_{1,2}$ are
\eq{
\gamma_1 &= \frac{\gamma_2\, \ov\tau_s\, \bigl(a_s (3 + 4 a_w \ov\tau_s)-4
  a_w\bigr)}{a_w(-1 + 4 a_s \ov\tau_s)}\,, \\[0.2cm]
\gamma_2&=  \frac{24 {\sqrt 2}\, A_w \, a_w\, \xi_s \, |W_0|^2 \, {\sqrt{\ov\tau_s}}(-1+a_s \ov\tau_s) }{a_s \,
\ov{\cal V}^3 \, (1-4 a_s \ov\tau_s) }\, .
}
The minimization condition for $\tau_w$  boils down to
\begin{subequations}
\bea
\label{eq:Tau8}
& & \partial_{\tau_w} V({\tau_w})\vert_{\ov\tau_w} = 0 \Rightarrow a_w \ov\tau_w = 1 + \frac{\ov\tau_s \bigl(a_s (3 + 4 a_w \ov\tau_s)-4 a_w\bigr)}{1 - 4 a_s \ov\tau_s}\,,
\eea
\bea
\label{eq:del2VTau8}
& & \partial^2_{\tau_w} V({\tau_w})|_{\ov\tau_w} = \frac{24\sqrt 2 a_w^2 \, A_w \, \xi_s\, |W_0|^2 \,
\sqrt{\ov\tau_s} \, (-1 + a_s \ov\tau_s) \, e^{-a_w \ov\tau_w}}{a_s \, \ov{\cal V}^3 \, (-1 +4 a_s \ov\tau_s)} >0\,.
\eea
\end{subequations}
Now we observe that, for the natural values
$a_s = 2 \pi = a_w$ and $\ov\tau_s>1$, one obtains  $\ov\tau_w<0$.
Even though, for $a_s = \frac{2\pi}{n_a}, a_w = 2\pi$ with
appropriate value of  $n_a$ and a marginally trustable value of $\ov\tau_s$,
it might be possible to get a positive value of
$\ov\tau_w$, it turns out that still $\ov\tau_w < 1$.
Thus, we conclude that in this scheme
$\tau_w$ gets stabilized outside the range of validity of the leading order
low-energy effective action. Therefore, this minimum is not trustable.

\subsubsection*{Scheme 2: racetrack}

In this section, we will perform a similar analysis
for a racetrack type extension of the superpotential.
The intention is that this introduces more parameters into the problem,
increasing the chances that, for certain choice(s) of a set of parameters,
trustable minima for the volume of the Wilson line divisor
can be found.
The poly-instanton corrected racetrack superpotential takes the form
\eq{
\label{eq:KW+race}
W = W_0 + A_s \, e^{-a_s T_s}+ A_s A_w \, e^{-a_s T_s - a_w T_w} - B_s \,
e^{-b_s T_s}- B_s B_w \, e^{-b_s T_s - b_w T_w} \, .
}
The generic form of the induced scalar potential
can be found in eq.\eqref{eq:Vgen+race} in appendix \ref{appendixscalar}.
Now, we proceed with the same two step strategy for studying K\"ahler
moduli stabilization as done in the previous section.

In the large volume limit, the most dominant contributions to the
generic scalar potential ${\bf V}(\tau_b,\tau_s,\tau_w,\rho_s,\rho_w)$
are again collected by three types of terms.
In the absence of poly-instanton effects, they  simplify to\footnote{We use bold font notation for the scalar potential in case of
  racetrack superpotential.}
\eq{
{\bf V}({\cal V},\tau_s,\rho_s) \simeq {\bf V}_{\alpha^\prime}({\cal V}) +
{\bf V}_{\rm np1} ({\cal V},\tau_s,\rho_s) + {\bf V}_{\rm np2}({\cal V},\tau_s,\rho_s)\,,
}
where
\bea
\label{eq:Vgen+racelvs}
& & \hskip -0.8cm {\bf V}_{\alpha^\prime}  = \frac{3 {\, {\cal
      C}_{\alpha^\prime}} \,|W_0|^2}{2 \, {\cal V}^3}\, ,\nonumber\\[0.1cm]
& & \hskip -0.8cm {\bf V}_{\rm np1} =\frac{4\, a_s \, {A_s}\, W_0\, e^{-\, a_s {\tau_s}} \,  {\tau_s} \cos(\, a_s {\rho_s})}{{\cal V}^2}
 -\frac{4\, b_s \, {B_s}\, W_0\,  e^{-\, b_s {\tau_s}} \,  {\tau_s} \cos (\, b_s {\rho_s})}{{\cal V}^2}\,,\\[0.1cm]
& & \hskip -0.8cm {\bf V}_{\rm np2} = \frac{2\sqrt2\, a_s^2 \, A_s^2 \, e^{-2 \, a_s {\tau_s}} \sqrt{{\tau_s}}}{3\,  \xi_s \, {\cal V}}-\frac{4 \sqrt2 \,
a_s \, b_s \, {A_s}\, {B_s}\, e^{-(\, a_s+\, b_s) {\tau_s}} \sqrt{{\tau_s}} \cos \bigl((\, a_s-\, b_s) {\rho_s}\bigr)}{3\, \xi_s \, {\cal V}}\nonumber\\
& &  \hskip 0.6cm +\frac{2 \sqrt2\, b_s^2 \, B_s^2\, e^{-2 \, b_s {\tau_s}}
   \sqrt{{\tau_s}}}{3\, \xi_s \, {\cal V}}\, .\nonumber
\eea
The above scalar potential reduces to \eqref{eq:V_lvs} for $B_s = 0$.
Moreover, as expected, the potential \eqref{eq:Vgen+racelvs} does not
depend on the Wilson line  divisor volume modulus $\tau_w$.
Therefore, at this stage it  remains a flat-direction to be lifted
via sub-dominant poly-instanton effects.

The extremality conditions $\partial_{{\cal V}} {\bf V} =  \partial_{{\tau_s}} {\bf V}=\partial_{{\rho_s}} {\bf V}=0$ are  collectively given as
\eq{
\label{eq:Extremize}
W_0 &= \frac{\ov{\cal V} \,(b_s \ov\lambda_2 -a_s \ov\lambda_1) \, \Bigl[b_s \ov\lambda_2 (-1 + 4 b_s \ov\tau_s)
-a_s \ov\lambda_1 (-1 + 4 a_s \ov \tau_s)\Bigr]}{6 \sqrt2 \, \xi_s \, \sqrt{\ov\tau_s} \Bigl[b_s \ov\lambda_2 (-1 + b_s \ov\tau_s)-a_s \ov\lambda_1 (-1 + a_s \ov\tau_s)\Bigr]}\,,\\[0.2cm]
{\cal C}_{\alpha^\prime} &= \frac{32 \sqrt{2} \, \xi_s \, \ov\tau_s^{\frac{5}{2}} (b_s^2\, \ov\lambda_2 -
a_s^2\, \ov\lambda_1) \, \Bigl[b_s \ov\lambda_2 (-1 + b_s \ov\tau_s)-a_s \ov\lambda_1 (-1 + a_s \ov\tau_s)\Bigr]}{\Bigl[ a_s \ov\lambda_1(-1 + 4 a_s \ov\tau_s)
- b_s \ov\lambda_2(-1 + 4 b_s \ov\tau_s) \Bigr]^2} \, ,\\[0.2cm]
 a_s\, \ov\rho_s &= N \pi \quad   {\rm with} \quad
\ov\lambda_1 = A_s\, e^{-a_s \ov\tau_s} \quad \text{~and~} \ \  \ov\lambda_2 = B_s\, e^{-b_s
  \ov\tau_s}\; .
}
One finds that $\tau_s$ gets stabilized in terms of
$C_{\alpha^\prime}$ as $\ov\tau_s \sim ({C_{\alpha^\prime}})^{\frac{2}{3}}$
and then ${\cal V}$ gets stabilized via an exponential term $\exp(a_s\tau_s)$
(encoded in $\lambda_i$'s) so that  the overall volume of the
Calabi-Yau threefold is exponentially large. The intersection of the above three hypersurfaces in the moduli space of
$\{\tau_b,\tau_s,\rho_s\}$ determines the critical point(s) of the scalar potential.
These equations are hard to solve analytically, however a numerical analysis
gives the behavior of the scalar potential as shown in figure \ref{S12}.

Now, in addition to the leading order standard racetrack terms
\eqref{eq:Vgen+racelvs}, the generic potential \eqref{eq:Vgen+race}
involves sub-dominant contributions,  which are further suppressed by
powers of $\exp(-a_w \tau_w)$.
Collecting these sub-leading terms, in the single field approximation, the effective potential for $\tau_w$ becomes
\eq{
{\bf V}({\cal V},\tau_s,\tau_w,\rho_s,\rho_w) = {\bf V}^{\rm LVS}({\cal V},\tau_s,\rho_s)+
 {\bf V}(\tau_w,\rho_w) \,,
}
where, ${\bf V}^{\rm LVS}({\cal V},\tau_s,\rho_s)$ is the racetrack version of the standard large volume potential which stabilizes the K\"ahler  moduli $\tau_b$ (or ${\cal V}$) and $\tau_s$  at order ${\cal V}^{-3}$ and 
\eq{
 {\bf V}(\tau_w,\rho_w) &\sim \frac{4 W_0}{\ov{\cal V}^2}\Bigl[ \ov\lambda_1 \,
{A_w}\, e^{-\, {a_w} {\tau_w}} \, (a_s \ov\tau_s + a_w \tau_w)\, \cos( {a_w}
{\rho_w}) \\
&\phantom{aaaaaaaaaaaaaa} -\ov\lambda_2 \, {B_w}\, e^{- {b_w} {\tau_w}} \, (b_s {\ov\tau_s} + b_w \tau_w)\, \cos({b_w} {\rho_w})\Bigr]\\
&+\frac{4\sqrt 2\, (b_s \ov{\lambda}_2 -a_s \ov{\lambda}_1) \, \sqrt{\ov\tau_s}}{3\,\xi_s \,\ov{\cal V}}
\Bigl[ \ov\lambda_2 (b_s -b_w) B_w \, e^{- b_w \tau_w}
  \cos (b_w \rho_w) \\
 & \phantom{aaaaaaaaaaaaaaaaaaaaa} -\ov\lambda_1 (a_s -a_w) A_w \, e^{- a_w \tau_w} \, \cos(a_w \rho_w) \Bigr]\, \\
 &  \sim {\cal O}\left({\textstyle \frac{1}{{\cal
       V}^{3+p}}}\right).
}
Here, $p$ is a positive constant and, similar to \cite{Cicoli:2011ct}, it  is defined  such that the poly-instanton
contribution to the scalar potential adds an extra multiplicative factor $\left(\frac{1}{{\cal
    V}^{p}}\right)$ to the scalar potential as compared to the one  in the absence of poly-instanton effects.

After stabilizing the $\rho_w$-axion at its minimum and
using $a_w = b_w$ along with eliminating $W_0$ via the first relation
in eq.\eqref{eq:Extremize},
the above effective scalar potential for the  $\tau_w$ modulus can again be written in the simple form
\bea
\label{eq:Vtau8race}
& & {\bf V}(\tau_w) = V_0 + e^{-a_w \tau_w}\left(\mu_1 + \mu_2 \tau_w \right)\,.
\eea
Here $V_0, \mu_1, \mu_2$ are constants depending on the stabilized values of the heavier moduli as
\eq{
\label{eq:Vtau8}
\mu_1 &= \mu_0 \biggl[4 \ov\tau_s \bigl( (a_s -a_w) A_w \ov\lambda_1 -(b_s -a_w) B_w \ov\lambda_2\bigr)\\
 &+\frac{\ov\tau_s\, (b_s B_w \ov\lambda_2-a_s A_w \ov\lambda_1)
\,\bigl(a_s \ov\lambda_1 (-1+4a_s \ov\tau_s)-b_s \ov\lambda_2 (-1+4b_s
  \ov\tau_s)\bigr)
}{a_s \ov\lambda_1 (-1+a_s \ov\tau_s)-b_s \ov\lambda_2 (-1+b_s \ov\tau_s)}
\biggr]\,,\\[0.2cm]
\mu_2 &= \mu_0 a_w \left[\frac{\bigl( B_w \ov\lambda_2- A_w \ov\lambda_1\bigr) \bigl(a_s \ov\lambda_1 (-1+4a_s \ov\tau_s)
-b_s \ov\lambda_2 (-1+4b_s \ov\tau_s)\bigr)}{a_s \ov\lambda_1 (-1+a_s
    \ov\tau_s)-b_s \ov\lambda_2 (-1+b_s \ov\tau_s)}\right]
}
with
\eq{
\mu_0 = \frac{\sqrt2 (\, a_s \ov\lambda_1-b_s \ov\lambda_2)}{3 \, \xi_s
  \,\ov{\cal V} \, \sqrt{\ov\tau_s}}\, .
}
The extremality condition $\partial_{\tau_w} {\bf V}(\tau_w)|_{\ov\tau_w} = 0$
boils down to
\begin{subequations}
\bea
\label{eq:Tau8race}
& & a_w \ov\tau_w = 1-a_w\frac{\mu_1}{\mu_2}\,,
\eea
\bea
\label{eq:der2VtauRace}
\partial^2_{\tau_w} {\bf V}(\tau_w)|_{\ov\tau_w} = -a_w \mu_2 \,
\exp\Bigl(-1+\frac{a_w \,\mu_1}{\mu_2}\Bigr) > 0\; .
\eea
\end{subequations}
Note that, if we switch off the racetrack form in the superpotential,
the $\mu_i$ simplify and \eqref{eq:Tau8race} reduces  to
the relation \eqref{eq:Tau8} derived in the first scheme.
Recall that there we were facing the problem
of stabilizing $\tau_w$ such that we can trust the underlying effective
supergravity theory.

The sole effect of including  the  racetrack term is that the
parametrization of the $\mu_i$ change.
Tuning the model dependent parameters makes it now possible to stabilize
the modulus of the  Wilson line  divisor such that $\ov\tau_w>1$.
Demanding that this critical point is a minimum of the effective potential,
\eqref{eq:der2VtauRace} implies that the parameters
are to be tuned such that $\mu_2$ is negative. It follows that
$\mu_1$ has to be  positive, as  ${\mu_1/\mu_2} <0$ is required for
stabilizing $\tau_w$ inside the K\"ahler cone.

Furthermore, the parameters $\mu_i$ appearing in \eqref{eq:Vtau8race}
are corrected by sub-leading terms, which are suppressed by one more 
volume factor and hence do
not induce a significant shift in the stabilized modulus $\tau_w$. Also, it is important to emphasize that the form of the potential
\eqref{eq:Vtau8race} does not change via sub-leading contributions and only $V_0,~\mu_1$ and $\mu_2$ are corrected.
The  analytic studies done so far are confirmed
by the explicit numerical analysis
to be  presented in the upcoming paragraph.

\subsubsection*{Benchmark models}

\begin{figure}[!ht]
\centering
\includegraphics[scale=0.25]{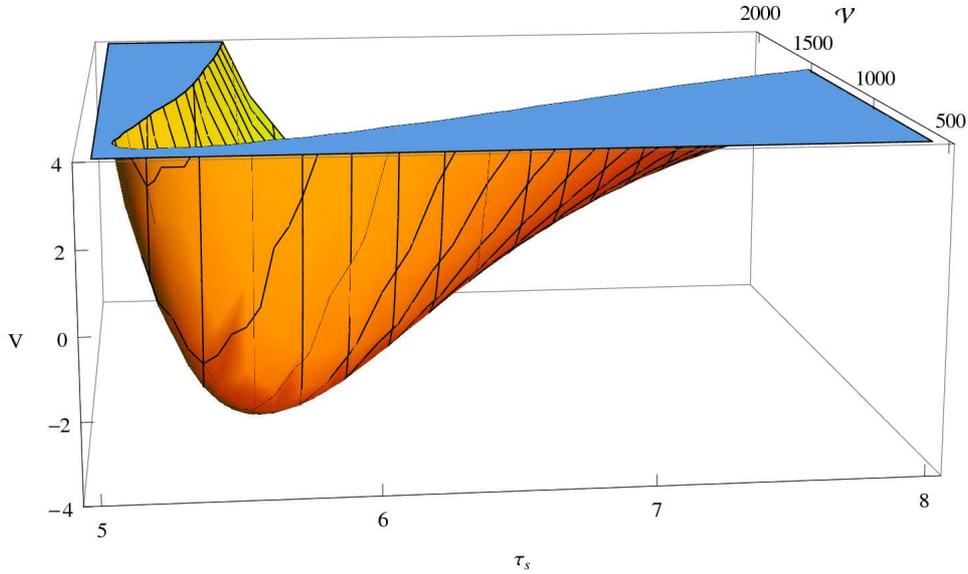}
\caption{The scalar potential $V({\cal V},\tau_s)$ (multiplied by $10^{7}$)
  vs ${\cal V}$ and $\tau_s$  for benchmark model ${\cal B}_1$ in the absence of poly-instanton effects.}\label{S12}
\end{figure}

\begin{figure}[!ht]
\centering
\includegraphics[scale=0.9]{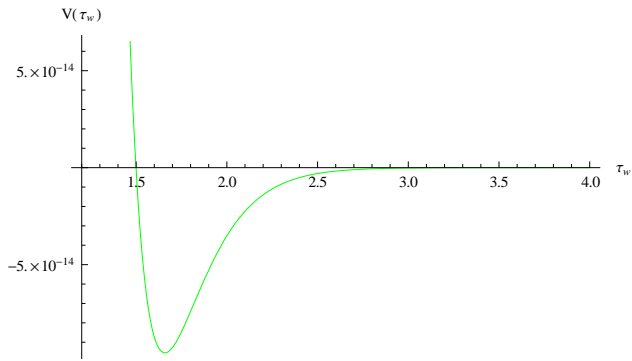}
\caption{The effective scalar potential $V(\tau_w)$  vs $\tau_w$ at stabilized value of the heavier moduli
 $\ov{\cal V}, \ov\tau_s$ for benchmark model ${\cal B}_1$.}\label{S17}
\end{figure}

In table \ref{table+parameter_choices} we present
the parameters for four benchmark models, where we have tuned
the parameters such that the subsequent stabilized values for divisor volumes
are within the K\"ahler cone and that we can trust
the supergravity theory.
\begin{table}[!ht]
\hspace{-0.4cm}
  \begin{tabular}{|c||cccccccc|cccc|}
\hline
&&&&&&&&&&&& \\[-0.2cm]
      & $A_s$ & $B_s$ & $A_w$ & $B_w$ & $a_s$ & $b_s$ &$g_s$ &$W_0$& $\mu_1$ & $\frac{\mu_2}{\mu_1}$&$\frac{V_0}{\mu_1}$& $p$\\[-0.2cm]
   &&&&&&&&&&&& \\[-0.2cm]
    \hline
    \hline
    &&&&&&&&&&&& \\[-0.2cm]
${\cal B}_1$ & 3&2& 0.5& 1.749&$\frac{2\pi}{7}$& $\frac{2\pi}{6}$ &0.12& -20 &$3.0\times10^{-8}$ & $-0.7$&$-11.4$&1.84\\[-0.2cm]
   &&&&&&&&& &&&\\[-0.2cm]
${\cal B}_2$ & 6&0.5& 0.1&4.771&$\frac{2\pi}{6}$& $\frac{2\pi}{5}$ &0.10& -10&$5.8\times10^{-10}$ & $-1.0$&$-117$&1.72\\[-0.2cm]
   &&&&&&&&& &&&\\[-0.2cm]
${\cal B}_3$ &8&0.8& 0.1& 6.5075&$\frac{2\pi}{5}$& $\frac{2\pi}{4}$ &0.11& -10&$5.1\times10^{-10}$ & $-1.0$& $-85.0$&1.61\\[-0.2cm]
   &&&&&&&&& &&&\\[-0.2cm]
${\cal B}_4$ &8 &0.8& 0.1& 17.143&$\frac{2\pi}{4}$& $\frac{2\pi}{3}$ &0.12& -5& $1.1\times10^{-10}$ & $-1.0$&$-62.1$&1.52
\\[-0.3cm]
   &&&&&&&&& &&&\\
     \hline
  \end{tabular}
  \caption{Parameters of the benchmark models, where
  the other model dependent parameters  are given as $\chi({\cal M}) = -136$,
$\xi_b = \frac{1}{36},~\xi_s  = \xi_{sw} =\frac{1}{6\sqrt2},~a_w = 2 \pi = b_w,$ and $\rho_s=0=\rho_w$.}
  \label{table+parameter_choices}
\end{table}
\noindent
In table \ref{table+stabilized+moduli} we list the values
of the moduli in the minimum.
\begin{table}[!ht]
  \centering
  \begin{tabular}{|c||ccc|}
\hline
&&& \\[-0.3cm]
    & \hskip 1cm $\ov\tau_s$ & \hskip 1cm $\ov\tau_w$ & \hskip 1cm $\ov{\cal V}$  \\[0.2cm]
    \hline
    \hline
    &&& \\[-0.3cm]
${\cal B}_1$ &\hskip 1cm 5.684& \hskip 1cm 1.658& \hskip 1cm 905.1\\[0.2cm]
${\cal B}_2$ &\hskip 1cm 6.607& \hskip 1cm 1.129& \hskip 1cm 941.0\\[0.2cm]
${\cal B}_3$ &\hskip 1cm 5.977& \hskip 1cm 1.161& \hskip 1cm 1015.3\\[0.2cm]
${\cal B}_4$ &\hskip 1cm 5.440& \hskip 1cm 1.124& \hskip 1cm 1093.2\\[0.2cm]
\hline
  \end{tabular}
  \caption{The stabilized values of divisor volume moduli corresponding to the respective benchmark
    models.}
  \label{table+stabilized+moduli}
\end{table}
For the benchmark model ${\cal B}_1$, the form of the scalar potential
$V({\cal V},\tau_s)$ as a function of  ${\cal V}$ and $\tau_s$ is
shown in figure \ref{S12}.
For fixed values
of the overall volume and of the volume of the small cycle $\tau_s$,
the dependence of the scalar potential on $\tau_w$
is shown in figure \ref{S17}.
The respective two plots for the
other benchmark models look quite similar.

\newpage
\section{Inflationary cosmology}
\label{sec_Inflationary Cosmology}

In this section we investigate whether the Wilson-line modulus $\tau_w$ can serve
as an inflaton field.
Since we are working in the framework of the LARGE volume scenario,
all K\"ahler moduli except the overall volume mode are potential  inflaton candidates.
The reason is that the ${\cal N}=1$ F-term scalar potential possesses the so-called
no-scale structure at tree level, which due to the   perturbative $\alpha^\prime$-correction
to the K\"ahler potential is broken only for the overall volume modulus.
Furthermore, the expansion of the dangerous prefactor $\exp(K)$ in the  F-term scalar potential
(which induces higher-dimensional operators and therefore corrections to $\eta$)
depends only on the overall volume of the  Calabi-Yau.
Therefore, in the absence of non-perturbative corrections (along with string-loop effects),
all K\"ahler moduli except the one for the overall volume of the Calabi-Yau remain flat
and hence are promising inflaton candidates.

Let us  explore the possibility of satisfying the  slow-roll
conditions for the resulting  effective potential of the $\tau_w$ modulus.
This is done by
investigating the dynamics of the lightest modulus $\tau_w$ while keeping the
heavier ones at their respective minimum. A detailed determination of the
moduli masses is presented in appendix \ref{appendixMass} where we find
the following  estimates \footnote{Note that
the mass of the lightest modulus in (\ref{eq:Mass}) has a different volume
  scaling than found in \cite{Cicoli:2011ct} where it was
  ${\cal V}^{-(3+p)/2} M_p$. The reason is simply the fact that the inflaton
  candidate in \cite{Cicoli:2011ct} is the fiber-volume of a K3-fibration
   and therefore coupled to the overall volume
  while for our case it is coupled to a blow-up mode. This is also
  reflected in the volume forms.}:
\bea
\label{eq:Mass}
& & \hskip -0.9cm M_{\tau_b} \sim {\cal O}(1) \, \frac{M_p}{{\cal V}^{\frac{3}{2}}}, \quad M_{\tau_s} \sim  {\cal O}(1) \, \frac{ M_p}{{\cal V}}\,, \quad M_{\tau_w} \sim  {\cal O}(1) \, \frac{M_p}{{\cal V}^{\frac{2+p}{2}}}\,.
\eea
From the parametrizations of our benchmark models ${\cal
  B}_i$, we find  that the value of $p$ lies in the range $1.5<p<1.8$
ensuring that $\tau_w$ is indeed the lightest modulus at the minimum. Furthermore, its mass is more exponentially suppressed in inflationary regime away from the minimum. Thus, a single field description can be
argued to be valid as the aforementioned mass-hierarchy is maintained while
the lightest modulus moves away from its  minimum.

Now, assuming that the heavier moduli are stabilized at their respective minimum position, we consider the effective
single modulus potential for $\tau_w$ modulus given by
\eqref{eq:Vtau8race}.
Furthermore, we assume that a suitable up-lifting of the $AdS$ minimum
to a $dS$ minimum can be processed via an appropriate mechanism (e.g.~by the introduction of anti-D3-branes).
The up-lifting term $V_{\rm up}$ needs to be such that the
up-lifted scalar potential acquires a small positive value (to be matched with
the cosmological constant) when all the moduli sit at their respective
minima. Thus, the resulting inflationary potential
looks like
\bea
& & V_{\rm inf} = V_{\rm up} + V_0 + e^{-a_w \tau_w}\left(\mu_1 + \mu_2 \tau_w \right)\,.
\eea

Now we consider a
shift in the volume modulus of the Wilson line divisor $\tau_w$ from
its minimum. Defining   $\hat{\tau}_w\equiv \tau_w-\ov\tau_w$ and
requiring that,
at the minimum, the small value of the cosmological constant is achieved via
adjusting the parameters appearing in the uplifting term such that $V_{\rm up} \simeq
-\left(V_0 + e^{-a_w \ov\tau_w}(\mu_1 + \mu_2 \ov\tau_w)\right)$, the
inflationary potential reduces to the following form\footnote{The factor
  $\left(\frac{g_s}{8 \pi}\right) e^{K_{\rm CS}}$ has been included for
  appropriate normalization of the scalar potential in Einstein  frame
  \cite{Conlon:2005jm,Conlon:2005ki}.
Throughout this section, we assume that $e^{K_{\rm CS}} \sim {\cal O}(1)$.},
\bea
\label{eq:Vinf}
& & V_{\rm inf} \simeq - \left(\frac{g_s}{8 \pi}\right)\, e^{K_{\rm CS}} \, \left[\frac{\mu_2\,  e^{-a_w \ov\tau_w}}{a_w} \, \Bigl(1-\left(1+ a_w \, \hat\tau_w\right) e^{-a_w {\hat{\tau}_w}}\Bigr)\right]\,.
\eea
Here $K_{\rm CS}$ denotes the K\"ahler potential for the complex structure
moduli.
Now, from the estimates in appendix \ref{appendixCanonical}, the leading order contributions to the canonical normalized fields $\chi_i$ are given as
\eq{
&  {\cal V} \sim \exp\Bigl({{\sqrt{{\textstyle \frac{3}{2}}}}\, \chi_1} \Bigr) \, , \qquad
\tau_s \sim \frac{1}{\lambda_s} \exp\Bigl({\sqrt{{\textstyle \frac{2}{3}}}}\, \chi_1\Bigr) \, \chi_2^{\frac 4 3} \, , \\[0.1cm]
& \tau_w +\tau_s \sim \frac{1}{\lambda_{sw}} \exp\Bigl({{\textstyle \sqrt{\frac{2}{3}}}}\, \chi_1\Bigr) \, \chi_3^{\frac 4 3}\,,
}
where $\lambda_s = \frac{2^{\frac 7 3} \xi_s^{\frac 2 3}}{3^{\frac 2 3}}$ and
$\lambda_{sw} = \frac{2^{\frac 7 3} \xi_{sw}^{\frac 2 3}}{3^{\frac 2 3}}$ are
constants which can be directly read off from the volume form. One  observes that the
canonically normalized moduli $\chi_2$ and $\chi_3$  are suppressed by a
large volume
factor ${\cal V}^{\frac 1 2}$ relative to  $\tau_s$ and $\tau_w$.
As a consequence, the
shift $\hat\chi_3$ in the canonically normalized inflaton candidate  $\chi_3$
will receive a sub-Planckian value so that  this
inflationary model lies in the class of small-field inflationary models.

\begin{figure}[!ht]
\centering
\includegraphics[scale=0.9]{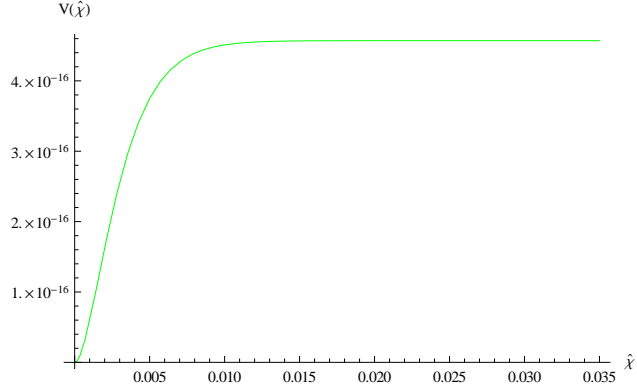}
\caption{The inflationary potential  $V(\hat\chi)$  vs the inflaton-shift
  $\hat\chi$ for the benchmark model ${\cal B}_1$ while keeping the heavier moduli at their respective minima
 $\ov{\cal V},~\ov\tau_s$. The flat region of the potential is the relevant inflationary region where slow-roll conditions are satisfied. }\label{S13}
\end{figure}

From now on, we drop the subscript for $\chi_3$ and denote the canonically
normalized inflaton field as $\chi$. The effective
inflationary potential \eqref{eq:Vinf}  can be expressed in terms of
the canonically normalized modulus $\chi$ as
\eq{
\label{eq:Vinfnew}
& \hskip -0.5cm V_{\rm inf} \simeq -\frac{g_s \, e^{K_{\rm CS}} \mu_2\,  e^{-a_w
    \ov\tau_w}}{ 8 \pi a_w} \, \Biggl[1-\biggl(1+ {\frac{\ov{{\cal V}}^{\frac 2
        3} \, a_w}{\lambda_{sw}} \, \Bigl({\chi}^{\frac 4 3} -
    {\ov\chi}^{\frac 4 3}\Bigr)}\biggr)\times \\
    &\phantom{aaaaaaaaaaaaaaaaaaaaaaaa}\, \exp\biggl({-{\frac{\ov{{\cal V}}^{\frac 2 3} \, a_w}{\lambda_{sw}} \, \Bigl({\chi}^{\frac 4 3} - {\ov\chi}^{\frac 4 3}\Bigr)}}\biggr)\Biggr]\,.
}
The form of this inflationary potential as a function of
the inflaton shift $\hat\chi$ is shown in  figure \ref{S13}.

\begin{figure}[!ht]
\centering
\includegraphics[scale=0.9]{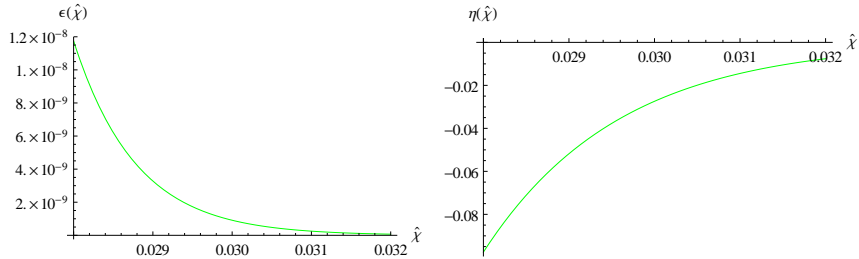}
\caption{For benchmark model ${\cal B}_1$, the two slow-roll parameters
$\epsilon(\hat\chi)$ and $\eta(\hat\chi)$ are
  displayed for inflaton shift $\hat\chi$ focussing on  locating the slow-roll regime where $\epsilon \ll |\eta| \ll 1$. }\label{S14}
\end{figure}

Sufficient conditions for realizing  slow-roll inflation are encoded in
two so-called slow-roll parameters defined as\footnote{The expressions for slow-roll parameters are defined in units
of reduced Planck mass ($M_{\rm pl}=1$) which is defined
as $M_{\rm pl} = \frac{1}{\sqrt{8\pi G}}$.}
\eq{
\epsilon(\chi) \equiv \frac{(\partial_{\chi} V_{\rm inf})^2}{2 V_{\rm inf}^2} \,, \, \, \qquad \eta(\chi) \equiv
\frac{\partial^2_{\chi} V_{\rm inf}}{V_{\rm inf}}\, .
}
For slow-roll inflation to occur, one needs
$\epsilon\ll1 \, , \, \eta\ll1$ in a region of the moduli space.  Our inflationary
potential \eqref{eq:Vinfnew} results in the following slow-roll parameters
\eq{
\label{eq:slow-roll}
 & \epsilon(\chi) = \frac{8 a_w^4 \, \ov{\cal V}^{\frac 8 3}\, \chi^{\frac 2 3} \bigl(\chi^{\frac 4 3}-\ov\chi^{\frac 4 3}\bigr)^2}{9 {\lambda_{sw}}^2 \biggl[{\lambda_{sw}}
 - {\lambda_{sw}}\exp\Bigl({\frac{a_w \ov{\cal V}^{\frac 2 3}\, (\chi^{\frac 4
       3}-\ov\chi^{\frac 4 3})}{{\lambda_{sw}}}}\Bigr) + a_w \ov{\cal V}^{\frac 2 3}\, (\chi^{\frac 4 3}
-\ov\chi^{\frac 4 3})\biggr]^2 }\,,\\[0.2cm]
 & \eta(\chi) = \frac{4 a_w^2 \, \ov{\cal V}^{\frac 4 3} \Bigl(4 a_w \ov{\cal V}^{\frac 2 3} \chi^{\frac 4 3} (\chi^{\frac 4 3}-{\ov\chi}^{\frac 4 3})
+{\lambda_{sw}}(-5 \chi^{\frac 4 3} + {\ov\chi}^{\frac 4 3})\Bigr)}{9 {\lambda_{sw}}^2
  \chi^{\frac 2 3} \biggl({\lambda_{sw}} - {\lambda_{sw}} \exp\Bigl(\frac{a_w \ov{\cal V}^{\frac 2 3}\,
(\chi^{\frac 4 3}-{\ov\chi}^{\frac 4 3})}{{\lambda_{sw}}}\Bigr)  + a_w \ov{\cal V}^{\frac 2 3}\, (\chi^{\frac 4 3}-{\ov\chi}^{\frac 4 3})\biggr)}\,.
}
These expressions contain exponentially suppressed large volume contributions
and hence there is a good chance to  satisfy the slow-roll
conditions. For appropriate shift of the inflaton from its minimum, we indeed
find slow-roll parameters satisfying $\epsilon \ll |\eta| \ll 1$.
The process of inflation starts from a point
$\chi^{\rm in}$ where $\epsilon\ll1$, i.e.~in a region where the potential is sufficiently flat.
This period of inflation comes to an end at a point $\chi^{\rm end}$
where the slow-roll parameter becomes $\epsilon \sim {\cal O}(1)$.
Figure \ref{S14} shows the variation of
slow-roll parameters w.r.t.\ the inflaton shift.

\subsubsection*{Numerical fitting and estimates of cosmological parameters}

Next, for  several cosmological parameters,  we compare  the numerical
estimates for  our model with  the experimental values extracted from the
data of the WMAP experiment measuring  the temperature  fluctuations
of the cosmological microwave background (CMB).
Let us start with one of the central cosmological observables,
the number of e-foldings during inflation.
This  can be computed as
\eq{
 N_e &\equiv \int_{\hat\chi^{\rm end}}^{\hat\chi^{\rm *}}
\frac{V_{\rm inf}}{(\partial_{\chi}V_{\rm inf})} \, d\hat\chi \\
  &\sim 3 {\lambda_{sw}}
\int_{\hat\chi^{\rm end}}^{\hat\chi^{\rm *}} \frac{\Bigl(-1 + {\lambda_{sw}} \exp\Bigl({\frac{a_w \ov{\cal
        V}^{\frac 2 3}\, (\chi^{\frac 4 3}-{\ov\chi}^{\frac 4 3})}{{\lambda_{sw}}}}\Bigr)
  \Bigr) + a_w \ov{\cal V}^{\frac 2 3} (-\chi^{\frac 4 3} + {\ov\chi}^{\frac 4
    3})}{4 a_w^2\, \ov{\cal V}^{\frac4 3} \, \chi^{\frac 1 3}\, (\chi^{\frac 4
    3}-{\ov\chi}^{\frac 4 3})}\,,
}
where $\hat\chi^{\rm *}$ is the inflaton shift at the horizon exit and
${\hat\chi^{\rm end}}$ the inflaton shift corresponding to a slow-roll
parameter $\epsilon =1$. For consistency with the observational data,
the above expression for the number of e-foldings has to be of
order $60$.
Moreover, in the large volume limit, we have the following relation
\eq{
\epsilon \ll |\eta| \sim \frac{1}{N_e}\, .
}

The other cosmological observables of interest are the amplitudes
and spectral indices for scalar and tensor perturbations.
These  are given (in $M_{\rm pl} =1$ units)  as
\eq{
\label{eq:cosmo+defs}
&  A_S = \frac{H}{2 \sqrt{2} \, \pi \sqrt{\epsilon}} \sim 5 \times 10^{-5} \, \, \, {{\rm ( \, imposed \, \, by \, \, COBE \, \, observations)}}\,,\\
& n_S = 1 + 2 \eta -6\epsilon  \, , \qquad \frac{d n_S}{d \ln k} = 16 \epsilon \eta -24 \epsilon^2 \, \, ,\\
 & A_T = \frac{\sqrt{2} H}{\pi} \,, \quad n_T = -2 \epsilon  \quad \text{and} \quad r \sim 12.4 \epsilon\;.
}
Here, $A_S,~A_T$ are amplitudes of scalar and tensor perturbations and $n_S,~n_T$ are their respective spectral indices. Furthermore, $r$ is the
tensor-to-scalar ratio, for which values $r>0.1$ can be
tested by the Planck satellite.

The dynamics of  the inflaton modulus $\chi$ as well as the Hubble parameter
$H$ is governed by the Friedman-equations (in units $M_{\rm pl} =1$)
\eq{
&  \frac{d^2 \chi}{d t^2} + 3 H \frac{d\chi}{d t} + \partial_{\chi} V_{\rm
    inf} = 0 \, , \\
&H^2  = \frac{1}{6} \left[2 V_{\rm inf} + \left(\frac{d\chi}{d t}\right)^2\right]\,.
}
In the slow-roll approximation, the Hubble parameter $H$ is such that $3 H^2 =
V_{\rm inf}$. Matching the COBE normalization for the density fluctuations
$\delta_H = 1.92 \times 10^{-5}$, results in a constraint defined in terms of a new parameter given as,
\eq{
{\cal A}_{\rm COBE} \equiv \left(\frac{g_s}{8 \pi}\right)\, e^{K_{\rm CS}}
\left(\frac{V_{\rm inf}^3}{(\partial_{\chi} V_{\rm inf})^2}\right) \sim  2.7
\times 10^{-7}\,.
}
Further, the scale of inflation can be estimated as
\eq{
M_{\rm inf} = V_{\rm inf}^{\frac{1}{4}} \,.
}
As, in the slow-roll regime, the inflationary potential is flat by definition,
the Hubble parameter and the scale of inflation (as defined above)  do
not vary significantly during the entire inflationary process. This is also
reflected in  figure \ref{S13}.

\begin{figure}[!ht]
\centering
\includegraphics[scale=0.9]{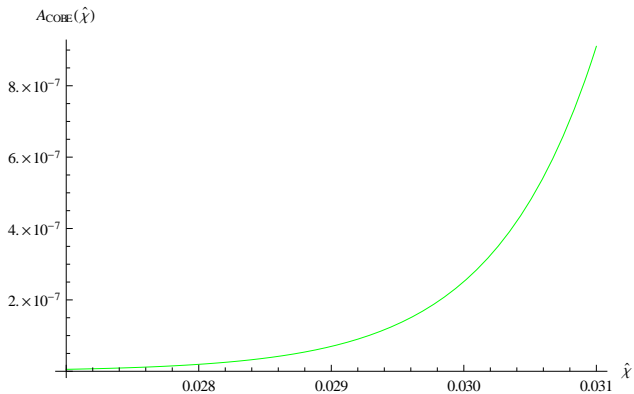}
\caption{The dependence of $A_{\rm COBE}$  on the inflaton shift $\hat\chi$
(for model ${\cal B}_1$).}\label{S15}
\end{figure}

\begin{figure}[!ht]
\centering
\includegraphics[scale=0.9]{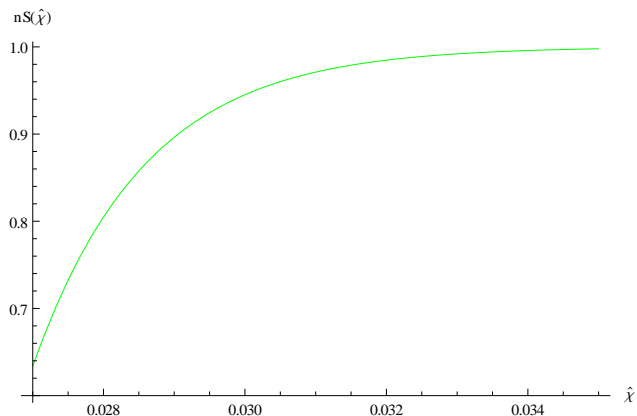}
\caption{The dependence of the spectral index $n_S$ on the
inflaton shift $\hat\chi$ (for model ${\cal B}_1$). }\label{S16}
\end{figure}

Now, we present the numerical estimates for these cosmological data for the benchmark models ${\cal B}_1-{\cal B}_4$ in the tables \ref{cosmo+parameters1} and  \ref{cosmo+parameters2}. Also, in the figures \ref{S15} and \ref{S16}, we show the dependence of
 $A_{\rm COBE}$ and $n_S$ on the value of the inflaton field.
\begin{table}[ht]
  \centering
  \begin{tabular}{|c||cc|ccc||c|}
\hline
&&&&&&\\[-0.3cm]
     & $\hat\chi^{\rm end}$ & $\hat\chi^{*}$ & $\epsilon$ & $\eta$ & $N_e$ &
$M_{\rm inf} ({\rm GeV})$ \\[0.2cm]
    \hline
    \hline
    &&&&&&\\[-0.3cm]
${\cal B}_1$ &$1.29\times 10^{-2}$& $3.08\times 10^{-2}$ &$3.2 \times 10^{-10}$& $-0.01644$ & $60.84$ &  $3.5\times 10^{14}$\\[0.2cm]
${\cal B}_2$ &$1.26\times 10^{-2}$& $3.00\times 10^{-2}$ &$2.9 \times 10^{-10}$& $-0.01607$ & $62.21$ &  $3.2\times 10^{14} $\\[0.2cm]
${\cal B}_3$ &$1.24\times 10^{-2}$& $2.95\times 10^{-2}$& $2.6\times10^{-10}$& $-0.01544$ & $64.79$  & $3.0 \times 10^{14}$ \\[0.2cm]
${\cal B}_4$ &$1.22\times 10^{-2}$& $2.90 \times 10^{-2}$ &$2.4\times 10^{-10}$& $-0.01532$ & $65.27$ & $2.2\times 10^{14}$ \\[0.2cm]
\hline
  \end{tabular}
  \caption{The numerical estimates for various cosmological observables -I.}
  \label{cosmo+parameters1}
\end{table}

\begin{table}[ht]
  \centering
  \begin{tabular}{|c|||ccc||ccc|}
\hline
&&&&&&\\[-0.3cm]
      & $n_S$ & ${\cal A}_{\rm COBE}$ & $r$  &$M_{\tau_s}$ & $M_{\cal V}$ & $M_{\tau_w}$ \\[-0.1cm]
   &&&& & {\tiny ($\times 10^{14}$ GeV) }& \\
    \hline
    \hline
    &&&&&&\\[-0.3cm]
${\cal B}_1$ &0.9671& $7.1\times 10^{-7}$ &$4.0\times 10^{-9}$ & $297$ & $3.49$& $0.28$\\[0.2cm]
${\cal B}_2$ &0.9679& $5.4\times 10^{-7}$ & $3.6\times 10^{-9}$ & $201$& $1.48$& $0.19$\\[0.2cm]
${\cal B}_3$ & 0.9691& $4.7\times 10^{-7}$ & $3.2\times 10^{-9}$ & $214$& $1.23$& $0.12$\\[0.2cm]
${\cal B}_4$  &0.9694& $1.5\times 10^{-7}$ & $3.0\times 10^{-9}$ & $120$ & $0.52$ & $0.03$\\[0.2cm]
\hline
  \end{tabular}
  \caption{The numerical estimates for various cosmological observables -II.}
  \label{cosmo+parameters2}
\end{table}

Note that, for benchmark model ${\cal B}_1$, the Hubble parameter
during the inflationary epoch takes the value
\eq{
H^2
= -\frac{g_s e^{K_{\rm CS}}}{8\pi } \left(\frac{\mu_2\,  e^{-a_w
    \ov\tau_w}}{a_w}\right)\sim 10^{-16}
}
and therefore  consistently reproduces the correct order of magnitude
for the scalar power spectrum
${\cal P}_S= |A_S|^2=H^2/(8\pi^2 \epsilon)$ with
the slow-roll parameter $\epsilon$ being of order $10^{-10}$.
Moreover, the  masses of the moduli $\tau_s, {\cal V}$ are
larger  than the value of the Hubble parameter ($H\sim 10^{10} $GeV)
along the entire inflationary trajectory
such that $M_{\tau_s} > M_{\cal V} > H \sim M_{\tau_w}$
and therefore, the heavier moduli remain  at their respective minimum
during the motion of the lightest (inflaton) modulus.
Thus, our assumption of single field inflation is
justified.

Let us close our numerical analysis with some general comments
on this inflationary scenario.
\begin{itemize}
\item{
Since the form of the inflationary
potential is exponentially flattened by the appearance of the large volume
$\ov{\cal V}$ factor, the respective cosmological observables  are
extremely sensitive to the inflaton shift. }
\item{ All four benchmark models
predict a negligible value of tensor-to-scalar ratio $r \sim 10^{-9}$ along
with negligible values for other tensor mode parameters $n_T,~A_T$.
This is something very common in inflationary
large volume models based on a blow-up modulus as inflaton field
\cite{Conlon:2005jm}.}
\item{Due to one of the inflationary
slow-roll parameter being very small ($\epsilon \sim 10^{-10}$), the running
of the scalar spectral index $\frac{d n_S}{d \ln k}$ is negligible.
}
\item{An equivalent convenient reformulation of the COBE normalization
of density perturbations $\delta_H = 1.92 \times 10^{-5}$ is given
as \cite{Conlon:2005jm}
\eq{
\left(\frac{V_{\rm inf}}{\epsilon}\right)^{\frac{1}{4}} = 6.6 \times 10^{16} \,
\, {\rm GeV}\, .
}
For instance, for benchmark model ${\cal B}_1$ we get
$8\times 10^{16}$ GeV. This estimate is
directly related to the scale of inflation, which
comes out as $M_{\rm inf} \sim 10^{14}\, {\rm GeV}$. Therefore, our
model belongs to the class of high scale inflationary models.}
\end{itemize}

\subsubsection*{Reheating}

Without going into too much detail of the post-inflationary processes,
let us comment on  the reheating of the universe  at the end of
inflation. Due to  possible couplings of the inflaton to both
observable and hidden sectors, its energy will be dumped
into various channels. Further, as argued in \cite{Cicoli:2010yj}, a mass gap can be generated via a dynamically generated scale $\Lambda$ appearing from the gaugino condensation in the field theroy supported on divisor $D_s$ which subsequently can make the hidden sector degrees of freedoms massive; to be more specific, the inflaton mass for our case is  lighter as $M_{\rm inf} \sim  \frac{M_p}{{\cal V}^{1+\frac{p}{2}}} < \Lambda \sim \frac{Mp}{{\cal V}^{5/6}}$ where $p>1$.  Hence, similar to \cite{Cicoli:2011ct}, the decay of the inflaton into this hidden sector particles is kinematically forbidden. 

In order for the reheating process not to interfere with the hot big bang
scenario and its successful description of nucleosynthesis,
it is important to determine the reheating temperature $T_{\rm rh}$.
This can be
computed from the decay width of the inflation into the various allowed
observable channels.
For this purpose, one has to compute the couplings of
the inflaton to the Standard Model gauge and matter fields.

For the string inspired model of interest  one proceeds as
follows \cite{Conlon:2007gk}.
Expanding the divisor volume moduli around their respective minima as $\tau_i
= \ov\tau_i + \delta\tau_i$, one arrives at the following form of a generic
Lagrangian,
\eq{
\label{Hannover96}
{\cal L} = \sum_{ij} K_{ij} (\partial_{\mu}\delta\tau_i) (\partial^{\mu}\delta\tau_j) -
\langle V\rangle -\frac{1}{2} \sum_{ij} V_{ij}\, \delta\tau_i\, \delta\tau_j + {\cal
  O}(\delta\tau^3)\, .
}
Let us denote by  $C_{ai}$ and $(M^2)_a$ the eigenvectors and eigenvalues
of the matrix  ${\cal M}_{ij}=\frac{1}{2} (K^{-1}\, V)_{ij}$.
Choosing the normalization $C^T_a\, K\, C_b=\delta_{ab}$
and redefining
$\delta\tau_i = \frac{1}{\sqrt2} C_{ai} \delta\phi_a$,
the Lagrangian \eqref{Hannover96} becomes
\eq{\
{\cal L} = \frac{1}{2} \sum_{a} (\partial_{\mu}\delta\phi_a)
(\partial^{\mu}\delta\phi_a) - \langle V\rangle -{\frac 1 2}\sum_a M_a^2 \,
\delta\phi_a^2 \,,
}
so that the $\delta\phi_a$ constitute a canonical normalized basis.
The details of this computation for our model
are  collected  in  appendix \ref{appendixMass}.

As we have not supported any (MS)SM-like model explicitly in the current
setup, we just focus on the inflaton coupling to  two light gauge-bosons living
on a  D7-brane worldvolume.
In a complete realistic model,  the  decay channels of
the moduli into matter fields have  to be analyzed, as well.
The relevant coupling arises from the gauge kinetic function and
gives the three-point coupling\footnote{A detailed systematic study of moduli
  dynamics and the computation of various couplings relevant for moduli
  thermalization can be
found in \cite{Anguelova:2009ht,Shukla:2010gv,Cicoli:2010ha}.}
\eq{
{\cal L}_{\rm gauge} = - \tau_i \, F_{\mu \nu}^{(i)}\, F^{\mu\nu}_{(i)}\, .
}
At the end of inflation, the inflaton  oscillates around its minimum value
so that, expanding $\tau_i=\delta\tau_i +\ov\tau_i$,
one gets  the couplings
\bea
& & {\cal L}_{\rm gauge} = - \frac{1}{4} G_{\mu \nu}^{(i)}\, G^{\mu\nu}_{(i)}
- \frac{\delta\tau_i}{4\ov\tau_i} \, G_{\mu \nu}^{(i)}\,  G^{\mu\nu}_{(i)}
\eea
for the canonically normalized field $G_{\mu\nu}^{(i)} = 2
F_{\mu\nu}^{(i)}\sqrt{\ov\tau_i}$.
As argued earlier, we can write $\delta\tau_i$ in terms of
fluctuations of canonically normalized field as given in appendix
\ref{appendixMass}.

To proceed, we assume now that the observable sector is localized on a stack of
$D7$-branes
wrapping the small divisor\footnote{This is not really realistic as we expect to encounter
the chirality problem described in \cite{Blumenhagen:2007sm}.} $D_s$. The numerical estimates for appropriate normalization 
for the corresponding divisor volume modulus fluctuation $\delta\tau_s$ which is relevant for the current purpose can be read from the appendix \ref{appendixMass}. For instance, the expression for benckmark model ${\cal B}_1$ from \eqref{eq:couplings} is,
\begin{eqnarray*}
& & \delta{\tau_s} = -0.258 \, \, \delta\chi_1 - 28.736 \, \,  \delta\chi_2 - 0.0168  \, \, \delta\chi_3\,.
\end{eqnarray*}
From above equation, one can read the volume scaling of
the coupling of the inflaton $\chi_3$ to  two  gauge bosons as  $ g_{\gamma\gamma\chi_3} \sim {0.0168}/({4 \bar\tau_s}) = 7.39 \times 10^{-4}\, \, M_p^{-1}$. Hence, using $M_{\chi_3} = 1.19 \times 10^{-5} \, \, M_p$, the decay width of the channel
$\chi_3\longrightarrow \{\gamma,\gamma\}$ is estimated from
\eq{
\Gamma_{\chi_3\rightarrow \{\gamma,\gamma\}} = \frac{N_g \,
  g_{\chi_3\gamma\gamma}^2 \, M_{\chi_3}^3}{64 \pi} 
}
Here, $N_g$ denotes the total number of gauge bosons (e.g.\ in MSSM it is 12) and subsequently, an estimated reheating temperature
is simply given as
\eq{
T_{\rm rh} \sim \sqrt{\Gamma M_{\rm pl}} \, .
}
which comes out to be $1.8 \times 10^7 \, \, {\rm GeV}$ for our benchmark models ${\cal B}_1$. In fact in all our benchmark models, the reheating temperature estimated via the aforementioend computations turns out to be
$T_{\rm rh}\sim  (10^6-10^7)\, {\rm GeV}$ which is much higher than the
big bang nucleosynthesis temperature $T_{\rm BBN}\simeq 1$ MeV.

\subsubsection*{Loop corrections}

Since the dynamics of the inflaton field is ruled by a non-perturbative
contribution to the superpotential, one must be worried about
more relevant corrections, i.e.~in particular string loop corrections
to the K\"ahler potential. Here one can distinguish between
open and closed string
loop corrections, i.e.~annulus and torus diagrams.
For moduli stabilization in the LARGE volume scenario, the open string
corrections have been shown to exhibit an
extended no-scale structure \cite{Berg:2007wt} implying  that their
contributions to the scalar potential are sub-dominant against the standard LARGE volume potential.

However, for inflation such corrections potentially induce an $\eta$-problem,
i.e.\ that, if the inflaton moves far away
from its minimum during inflation, loop-effects might be strong enough to
spoil the flatness of the potential \cite{Cicoli:2008gp,Cicoli:2010ha}.
Hence, similar to the case of blow-up inflation loop-effects might be crucial
for our case, as well.

What the effective theory on the tadpole canceling $D7$-brane configuration
is concerned, our model is clearly way too simple. In fact, we
only have a single stack of $D7$-branes right on top of the
$O7$-plane. These do wrap the small cycle, whose volume modulus
is $\tau_s$. Therefore, according to  \cite{Berg:2007wt}, the only
open string loop corrections come from Kaluza-Klein mode
exchange and are conjectured to lead to a contribution to the
K\"ahler potential of the form
\eq{
             \Delta K= \frac{\sqrt{\tau_s}\, {\cal E}(U,\ov U) }{ {\cal V}\,
               {\rm Im}(\tau)}\, .
}
Moreover, since the natural cut-off scale\footnote{This is different from
fiber inflation with poly-instantons \cite{Cicoli:2011ct},
where the cut-off scale was related  to  the size of
the base of the fibration, which explicitly depended on the inflaton
modulus.}
 of the effective four-dimensional theory
is $\Lambda=1/{\cal V}^{\frac 1 6}$,
we  expect the closed string loop correction to only depend
on the overall volume modulus ${\cal V}$.  Therefore, the loop corrections
do not explicitly depend on the inflaton $\tau_w$ (respectively the
canonically normalized field $\chi_3$)  so that, at least for
this very simple model, we do not expect an $\eta$-problem to be generated
by loop-effects.

\section{Conclusions}
\label{sec_Conclusions and Discussions}

In this article, we presented a new class of K\"ahler moduli inflation models
realized in the LARGE  volume scenario in the context of Type IIB orientifold
compactifications.  For prior work in this framework one could distinguish
essentially   two different
classes of K\"ahler moduli inflationary models, {\it blow-up inflation}
\cite{Conlon:2005jm} and {\it fiber inflation}
\cite{Cicoli:2008gp,Cicoli:2011ct}.
The first kind starts with a standard LARGE volume scenario and
considers one of the blow-up modes as the inflaton, whose leading order
contribution to the scalar potential arises  from a non-perturbative
instanton correction to the superpotential.
For the second class, the starting point is a Calabi-Yau threefold which
is $K3$-fibered and also contains shrinkable del Pezzo divisors.
The inflaton is given by the size modulus of the $K3$ fiber and
the potential can be generated either by loop \cite{Cicoli:2008gp} or by
poly-instanton effects \cite{Cicoli:2011ct}.

The model considered in this paper is different but shares some of
the features of both blow-up and fiber inflation.
The inflaton is neither a blow-up mode
nor a fiber mode but actually a K\"ahler modulus corresponding to a
so-called Wilson line divisor, which can be realized as a
$\mathbb P^1$ fibration over $\mathbb T^2$.
Like in fiber inflation, the leading order
contribution to the inflationary scalar potential arises from sub-leading (poly-instanton) corrections
to the superpotential, however, the contribution of the inflaton
to the Calabi-Yau volume form and therefore to the K\"ahler potential
is very similar to a blow-up mode.

As a prerequisite for inflation, in a simple representative model with three
K\"ahler moduli, we first studied the issue of moduli stabilization.
We considered two schemes.  The minimal setup contained just  the
leading order ${\alpha^\prime}$-correction to the K\"ahler potential as well
as a non-perturbative ($E3$ and poly-instanton) contribution to the
superpotential. Our conclusion was that it is not possible to stabilize
all the moduli within the geometric regime. In the second extended scheme
this could be improved by starting with a  racetrack form of  the
non-perturbative superpotential. Since this ansatz contains  more parameters,
all moduli were stabilized inside the geometric regime of the K\"ahler cone.

After identifying four benchmark models, we explored the possibility that
the volume of  Wilson-line divisor can play the role of  an inflaton field.
Being in a setup equipped with a swiss-cheese like volume form
and poly-instanton
effects, the inflationary signatures of the  model are  expected
to be a hybrid of the blow-up and
fiber-inflation models, i.e.~one expects to see features
from both inflationary scenarios.
Indeed, the intermediate steps in moduli stabilization process
and the form of inflationary potential as a function of the divisor volume moduli look quite
similar to those of \cite{Cicoli:2011ct}.
However, as the field redefinitions
related to canonical normalization of  kinetic terms
are  similar to those of the moduli involved in
blow-up inflationary models, the expressions for slow-roll parameters and
other cosmological observables turn out to be  rather similar to the blow-up case.

Due to the appearing exponential suppression factors in the large volume
regime, the slow-roll
conditions are easily satisfied. Moreover, for certain choices of the initial
parameters one can bring the number of
e-foldings, the density perturbation  amplitude and the spectral index
in perfect agreement with the current observational constraints.
The  model predicts a negligible value for the parameters involving tensor
perturbation modes, like  the
tensor-to-scalar ratio $r$, the spectral index $n_T$ and the amplitude $A_T$.
The inflaton rolls over sub-Planckian distances so that
the model belongs to the  class of `small' field inflation.
Furthermore, the scale of inflation could be shown to be quite high (of the
order $10^{14}$GeV), which means that we are dealing with
a  `high' scale inflationary model.

Contrary to  blow-up inflationary models,
the stabilized value of the Calabi-Yau volume turns out  to be relatively  low
(${\cal V}\sim {\cal O}(10^3)$) for generating sufficient amount of scalar perturbation along with ${\cal O}(60)$ number of e-foldings.
At the end of inflation, the decay of the inflaton into the degrees of freedom
of a (toy) visible sector cause a reheating temperature of the order $T_{\rm rh}\sim (10^6-10^7)$GeV which is much larger than the lower bound given by the
big bang nucleosynthesis temperature $T_{\rm BBN}\sim {\rm MeV}$. However, addressing this issue needs more attention in a less simple setup as compared to ours where one could explicitly distinguish the hidden and visible sectors to get the relevant couplings for estimating all decay channels. In such cases, there might be a possibility of inflaton dumping most of its energy to the hidden sector instead of visible sector\footnote{We thank A. Mazumdar for bringing our attention to \cite{Cicoli:2010yj} for such possibilities.}. 

In this paper we have only analyzed the  most simple conceivable set-up
for having a poly-instanton generated contribution to the superpotential.
Due to the small number of involved K\"ahler moduli, this clearly
does not admit enough freedom to also support a realistic (MS)SM on
intersecting, fluxed $D7$-branes. For a more thorough analysis of the
interactions between the inflaton sector and the visible sector
and of possibly dangerous loop-corrections
an extension of our model in this direction is necessary.

\subsubsection*{Acknowledgments}
We would like to thank Michele Cicoli,  Shanta de Alwis, Sebastian Halter, Stefan Hohenegger, 
Dieter L\"ust, Pablo Soler, Zhongliang Tuo and Angel Uranga for helpful discussions. 
RB is grateful to the Simons Center for Physics and Geometry at Stony Brook University for hospitality.
XG is supported by the MPG-CAS Joint Doctoral Promotion Programme. PS is supported by a postdoctoral research fellowship from the Alexander von Humboldt Foundation.

\begin{appendix}
\section{K\"ahler metric and its inverse}
\label{appendixInvK}
The K\"ahler potential is given as $K = - 2 \, \ln {\cal Y}$, where
${\cal Y}\equiv {\cal V}+C_{\alpha^\prime}$
is defined in terms of K\"ahler coordinates as
\eq{
{\cal Y}= \xi_b (T_b+\bar{T}_b)^{\frac{3}{2}}-\xi_s
(T_s+\bar{T}_s)^{\frac{3}{2}}-\xi_{sw}
\Bigl((T_s+\bar{T}_s) + (T_w+\bar{T}_w)\Bigr)^{\frac{3}{2}} +
C_{\alpha^\prime}\,,
}
and the perturbative ${\alpha^\prime}^3$ contribution is being given as
$C_{\alpha^\prime} = - \frac{\chi(CY_3) \, ({{\tau}-\bar\tau})^{\frac{3}{2}} \zeta[3] }{4 (2 \pi)^3 \,
({2i})^{\frac{3}{2}}}$. The expression of K\"{a}hler derivative can be read as
\bea
& & \partial_{T_b} K = -\frac{3 \sqrt{2} \, \xi_b \, \sqrt{\tau_b}}{{\cal Y}} \, \, , \nonumber\\
& & \partial_{T_s} K = \frac{3 \sqrt{2} \, (\xi_s \, \sqrt{\tau_s} + \xi_{sw} \, \sqrt{\tau_s +  \tau_w})}{{\cal Y}} \, \, ,\\ \, \,
& & \partial_{T_w} K = \frac{3 \sqrt{2} \, \xi_{sw} \, \sqrt{\tau_s + \tau_w}}{{\cal Y}}\,.\nonumber
\eea
The expressions for the various components of the K\"ahler metric are given as under,
\begin{eqnarray*}
\label{eq:Kmetric}
& & K_{T_b {\bar T}_b} = -\frac{{3 \sqrt{2} \xi_b \, {\cal Y}}-36 \xi_b^2 {\tau_b}^{3/2}}{4
   \, {\cal Y}^2 \,{\sqrt{{\tau_b}}}} \, \, , \nonumber\\
& & K_{T_b {\bar T}_s} =-\frac{9 \xi_b \sqrt{{\tau_b}} \left(\sqrt{{\tau_s}} \xi_s+ \xi_{sw}
   \sqrt{{\tau_s}+{\tau_w}}\right)}{\, {\cal Y}^2} = K_{T_s {\bar T}_b} \, , \nonumber\\
\end{eqnarray*}
\bea
& & K_{T_b {\bar T}_w} = -\frac{9 \xi_b \, \xi_{sw}\sqrt{{\tau_b}} \sqrt{{\tau_s}+{\tau_w}}}{\, {\cal Y}^2} = K_{T_w {\bar T}_b} \, , \\
& & K_{T_s {\bar T}_s} = \frac{3 \left(\sqrt2 {\cal Y}\, \left(\frac{\xi_s}{\sqrt{{\tau_s}}}
+\frac{\xi_{sw}}{\sqrt{{\tau_s}+{\tau_w}}}\right)+12\left(\xi_s \sqrt\tau_s + \xi_{sw} \sqrt{\tau_s+\tau_w}\right)^2\right)}{4 \, {\cal Y}^2}\, , \nonumber\\
& & K_{T_s {\bar T}_w} = \frac{3 \xi_{sw}\left(\sqrt2 {\cal Y} + 12 (\tau_s + \tau_w)\left(\xi_s\sqrt\tau_s
 + \xi_{sw} {\sqrt{\tau_s +\tau_w}}\right)\right)}{4 \, {\cal Y}^2 \sqrt{{\tau_s}+{\tau_w}}} =  K_{T_w {\bar T}_s} \, , \nonumber\\
 & & K_{T_w {\bar T}_w} = \frac{3 \xi_{sw}(\sqrt2 {\cal Y} + 12 \xi_{sw} ({{\tau_s}+{\tau_w}})^{3/2})}{4\, {\cal Y}^2 \sqrt{{\tau_s}+{\tau_w}}}\,.\nonumber
\eea
The expressions for various components of the inverse K\"ahler metric are given as under,
\bea
\label{eq:Kmetric2}
& & K^{T_b {\bar T}_b} = -\frac{2\sqrt2 \, {\cal Y} \, \sqrt\tau_b}{3\xi_b} + \frac{8 {\cal Y} \,
\tau_b^2}{{\cal Y} - 3 {\cal V}} \, \,, \, \,  K^{T_b {\bar T}_i} =-\frac{8 {\cal Y} \,
\tau_b \, \tau_i}{{\cal Y} - 3 {\cal V}} = K^{T_i {\bar T}_b} , \, \{i = s,w\} , \nonumber\\
& & K^{T_s {\bar T}_s} = \frac{2\sqrt2 \, {\cal Y} \, \sqrt\tau_s}{3\xi_s} - \frac{8 {\cal Y} \,
\tau_s^2}{{\cal Y} - 3 {\cal V}}\, , \, K^{T_s {\bar T}_w} = -\frac{2\sqrt2 \, {\cal Y} \,
\sqrt\tau_s}{3\xi_s} - \frac{8 {\cal Y} \, \tau_s \tau_w}{{\cal Y} - 3 {\cal V}} =  K^{T_w {\bar T}_s} \,,\nonumber\\
 & & K^{T_w {\bar T}_w} = \frac{2\sqrt2 \, {\cal Y} \, {\sqrt{\tau_s + \tau_w}}}{3\xi_{sw}}
- \frac{8 {\cal Y} \, \tau_w^2}{{\cal Y} - 3 {\cal V}}+\frac{2\sqrt2 \, {\cal Y} \, {\sqrt{\tau_s}}}{3\xi_s}\,,
\eea
where in terms of divisor volumes $\tau_i$'s, we have
\eq{
{\cal Y} \equiv {\cal V} + C_{\alpha^\prime}= 2\sqrt2\left(\xi_b \,
\tau_b^{3/2}- \xi_s \, \tau_s^{3/2}-\xi_{sw} \,
(\tau_s+\tau_w)^{3/2}\right)+C_{\alpha^\prime}\,.
}
The above form of the K\"ahler metric and its inverse are exact for the given ansatz of K\"ahler potential.
After neglecting the $\alpha^\prime$-corrections in the large volume limit (i.e. using  ${\cal Y} \sim {\cal V}\sim2\sqrt2 \, \xi_b \,
\tau_b^{3/2}$), this simplifies to the form
\bea
\label{eq:Ksimp}
\hskip-1cm K^{A\bar{B}} \sim \left(
\begin{array}{lll}
\frac{4 \tau_b^2}{3}& 4 \tau_b \tau_s   & 4 \tau_b \tau_w\\
4 \tau_b \tau_s & \frac{8 \xi_b \,  \tau_b^{3/2} \, \sqrt\tau_s}{3 \, \xi_s}   & -\frac{8 \xi_b \,  \tau_b^{3/2} \, \sqrt\tau_s}{3 \, \xi_b} \\
4 \tau_b \tau_w & -\frac{8 \xi_b \,  \tau_b^{3/2} \, \sqrt\tau_s}{3 \, \xi_s}  & \frac{8 \xi_b \,  \tau_b^{3/2} \, (\xi_s {\sqrt{\tau_s+\tau_w}} + \xi_{sw} \sqrt\tau_s)}{3 \,\xi_s \, \xi_{sw}}
\end{array}
\right)\,.
\eea
The above form reflects  the strong swiss-cheese nature of the volume form for
the $\{T_b,T_s\}$ moduli sector, while the $T_w$ modulus makes a (slight) difference.

\section{The F-term scalar potential}
\label{appendixscalar}

From the generic expressions of the K\"ahler potential $K$ and the superpotential
$W$, one can compute the scalar potential from \eqref{eq:Vgen}.

\subsubsection*{Scheme 1: minimal}

After collecting the most dominant contributions of three types of terms, the
generic expression of the F-term scalar potential is given as
\eq{
V(\tau_b,\tau_s,\tau_w,\rho_s,\rho_w) \equiv V({\cal
  V},\tau_s,\tau_w,\rho_s,\rho_w)= V_{\alpha^\prime} + V_{\rm np1} + V_{\rm np2}\,,
}
where
\begin{eqnarray}
& & V_{\alpha^\prime}  = \frac{3 {\, {\cal C}_{\alpha^\prime}} \,|W_0|^2}{2 {\cal V}^3}\,,\nonumber\\
& & V_{\rm np1} =\frac{4\, {W_0}}{{\cal V}^2}\biggl[{\,{A_s}} \,
 e^{-{a_s \tau_s}-{a_w \, \tau_w}} \biggl\{{\,a_s} e^{{a_w \, \tau_w}} {\tau_s} \cos ({\,a_s} {\rho_s})\\
 & & \hskip 1.7in +{\,{A_w}} ({a_s \tau_s}+{a_w \, \tau_w}) \cos ({\,a_s} {\rho_s}+{\,{a_w}} {\rho_w})\biggr\} \biggr]\,,\nonumber\\
& & V_{\rm np2} = \frac{2\sqrt2}{3 \, \xi_s \, \xi_{sw} \, {\cal V}} \biggl[{\,{A_s}}^2 e^{-2 ({a_s \tau_s}+{a_w \, \tau_w})}
\biggl(\xi_{sw} \left({\,{A_w}}^2+e^{2 {a_w \, \tau_w}}\right) \sqrt{{\tau_s}} {\,a_s}^2 \nonumber\\
& & \hskip 1cm +2 ({\,a_s}-{\,{a_w}}) {\,{A_w}}  \xi_{sw} \, e^{{a_w \, \tau_w}} \sqrt{{\tau_s}} \cos ({\,{a_w}} {\rho_w}) {\,a_s}-2 {\,{a_w}} {\,{A_w}}^2 \xi_{sw} \sqrt{{\tau_s}} {\,a_s}\nonumber\\
& & \hskip 2.1in +{\,{a_w}}^2 {\,{A_w}}^2 \left(\sqrt{{\tau_s}+{\tau_w}} \xi_s+ \xi_{sw} \sqrt{{\tau_s}}\right)\biggr)\biggr]\,.\nonumber
\end{eqnarray}

\subsubsection*{Scheme 2: racetrack}
After collecting the most dominant contributions of three types of terms,
the generic expression of the F-term scalar potential with the racetrack superpotential, is given as
\bea
\label{eq:Vgen+race}
& & {\bf V}(\tau_b,\tau_s,\tau_w,\rho_s,\rho_w) \equiv {\bf V}({\cal
  V},\tau_s,\tau_w,\rho_s,\rho_w) = {\bf V}_{\alpha^\prime} + {\bf V}_{\rm
  np1} + {\bf V}_{\rm np2}\,,
\eea
where
\begin{eqnarray}
& & \hskip-0.7cm {\bf V}_{\alpha^\prime}  = \frac{3 {\, {\cal C}_{\alpha^\prime}} \,|W_0|^2}{2 \, {\cal V}^3}\,,\nonumber\\
& & \hskip-0.7cm {\bf V}_{\rm np1} =\frac{4 \, {W_0}}{{\cal V}^2} \biggl\{\, a_s \, {A_s} e^{-\, a_s {\tau_s}}  {\tau_s} \cos(\, a_s {\rho_s})  -{\, b_s \, {B_s} e^{-\, b_s {\tau_s}} {\tau_s} \cos (\, b_s {\rho_s})} \\
& & \hskip 3cm +{\, {A_s} \, {A_w} e^{-\, a_s {\tau_s}-{a_w} {\tau_w}} (a_s {\tau_s}+a_w \tau_w) \cos(a_s {\rho_s}+{a_w} {\rho_w})} \nonumber\\
& & \hskip 4cm -{ {B_s} \, {B_w} e^{-\, b_s {\tau_s}-{a_w} {\tau_w}}  (b_s {\tau_s}+a_w \tau_w) \cos(b_s {\rho_s}+{a_w} {\rho_w})}\biggr\}\,,\nonumber\\
& & \hskip-0.7cm {\bf V}_{\rm np2} = \frac{2\sqrt2 \,\sqrt{{\tau_s}}}{3  \xi_s \, {\cal V}}\biggl\{\, a_s^2 \, {A_s}^2 \,e^{-2 \, a_s {\tau_s}} +{\, b_s^2 \, {B_s}^2 e^{-2 \, b_s {\tau_s}} } -{2 \, a_s \, b_s \, {A_s}\, {B_s} e^{-(\, a_s+\, b_s) {\tau_s}}} \nonumber\\
& & \hskip 0.5cm \times \cos((\, a_s-\, b_s) {\rho_s})\biggr\} + \frac{2 \sqrt{2} e^{-2 (a_s+b_s) {\tau_s}}}{3 \xi_s  \, \xi_{sw} \, {\cal V}} \biggl\{e^{-2 a_w {\tau_w}}
\bigl(A_s^2 a_w^2 \xi_s e^{2 b_s {\tau_s}} \sqrt{{\tau_s}+{\tau_w}}A_w^2\nonumber
\end{eqnarray}
\begin{eqnarray*}
& & \hskip 0.5cm +2 \,{A_s} \, B_s e^{(a_s+b_s) {\tau_s}} \bigl(-\, A_w \xi_s \, B_w \sqrt{{\tau_s}+{\tau_w}}
a_w^2-a_s b_s \, \xi_{sw} e^{2 a_w {\tau_w}} \sqrt{{\tau_s}}+(a_s-a_w) \nonumber\\
& & \hskip 0.5cm \times\, A_w  (a_w-b_s) \, B_w \xi_{sw} \, \sqrt{{\tau_s}}\bigr) \cos((a_s-b_s) {\rho_s})
-2 \xi_{sw} e^{a_w {\tau_w}} \sqrt{{\tau_s}} \bigl(\bigl(a_s (a_w-a_s) \,\nonumber\\
& & \hskip 0.5cm \times  A_w e^{2 b_s {\tau_s}} \,{A_s}^2 +(a_w-b_s) b_s \, B_s^2 \, B_w e^{2 a_s {\tau_s}}\bigr) \cos (a_w {\rho_w})+\,{A_s} \, B_s e^{(a_s+b_s) {\tau_s}} (a_s\nonumber\\
& & \hskip 0.5cm \times (b_s-a_w) \, B_w  \cos(a_s {\rho_s}-b_s {\rho_s}-a_w {\rho_w}) +(a_s-a_w) \, A_w b_s \cos (a_s {\rho_s}-b_s {\rho_s}\nonumber\\
& & \hskip 0.5cm +a_w {\rho_w}))\bigr)+\xi_{sw} e^{2 a_w {\tau_w}} \bigl(a_s^2 e^{2 b_s {\tau_s}} \,{A_s}^2 +b_s^2 \, B_s^2 e^{2 a_s {\tau_s}}\bigr) \sqrt{{\tau_s}}+\xi_{sw} \bigl(\,{A_s}^2 (a_s-a_w)^2 \nonumber\\
& & \hskip 0.5cm \times e^{2 b_s {\tau_s}} \, A_w^2+(a_w-b_s)^2 \, B_s^2  B_w^2 e^{2 a_s {\tau_s}}\bigr) \sqrt{{\tau_s}}+a_w^2 b \, B_s^2 \, B_w^2 e^{2 a_s {\tau_s}} \sqrt{{\tau_s}+{\tau_w}}\bigr)\biggr\}\,.\nonumber
\end{eqnarray*}

\section{Canonical normalization of K\"ahler moduli}
\label{appendixCanonical}
The simplified K\"ahler metric after neglecting the $\alpha^\prime$-correction in the large volume limit is given as
\begin{eqnarray*}
K_{I \bar{J}} = \left(
\begin{array}{lll}
 -\frac{\frac{3 \sqrt{2} \xi_b \, {\cal V}}{\sqrt{{\tau_b}}}-36 \xi_b^2 {\tau_b}}{4
   \, {\cal V}^2} & -\frac{9 \xi_b \sqrt{{\tau_b}} \left(\sqrt{{\tau_s}} \xi_s+ \xi_{sw}
   \sqrt{{\tau_s}+{\tau_w}}\right)}{\, {\cal V}^2}  & -\frac{9 \xi_b \,  \xi_{sw}\sqrt{{\tau_b}} \sqrt{{\tau_s}+{\tau_w}}}{\, {\cal V}^2} \\
-\frac{9 \xi_b \sqrt{{\tau_b}} \left(\sqrt{{\tau_s}}\xi_s + \xi_{sw} \sqrt{{\tau_s}+{\tau_w}}\right)}{\, {\cal V}^2} & \frac{3 \left(\frac{\xi_s}{\sqrt{{\tau_s}}}+\frac{\xi_{sw}}
{\sqrt{{\tau_s}+{\tau_w}}}\right)}{2 \sqrt{2} \, {\cal V}}  & \frac{3 \xi_{sw}}{2 \sqrt{2} \, {\cal V} \sqrt{{\tau_s}+{\tau_w}}} \\
  -\frac{9 \xi_b \xi_{sw} \sqrt{{\tau_b}} \sqrt{{\tau_s}+{\tau_w}}}{\, {\cal V}^2}
   & \frac{3 \xi_{sw}}{2 \sqrt2\, {\cal V} \sqrt{{\tau_s}+{\tau_w}}}  & \frac{3 \xi_{sw}}{2 \sqrt2\, {\cal V} \sqrt{{\tau_s}+{\tau_w}}}
\end{array}
\right)\,.\nonumber\\
\end{eqnarray*}
Utilizing the above K\"ahler metric, all kinetic terms for the respective
moduli can be written as
\eq{
{\cal L}_{\rm kinetic}({\cal V},\tau_s,\tau_w,\rho_b,\rho_s,\rho_w) \equiv
K_{I\bar J} (D_{\mu} T_I) ({\bar D}^\mu {\bar T}_{\bar J}) = \frac{1}{2}
\sum_{i=1}^6 ({\partial_\mu} {\chi_i})({\partial^\mu} {\chi_i})\,,
}
where the summation is over six real moduli $\chi_i$ which constitute the canonically normalized forms of six real moduli $\{{\cal V},\tau_s,\tau_w,\rho_b,\rho_s,\rho_w\}$. 
Furthermore, as for the time being  we are not interested in the dynamics of
axion moduli, we consider the terms corresponding to the (divisor) volume
modulus only. The leading and next-to-leading order kinetic terms for the volume moduli $\{{\cal V},\tau_s,\tau_{w}\}$ can be easily read off as
\bea
& & {\cal L}_{\rm kinetic}({\cal V},\tau_s,\tau_{sw})\equiv \frac{1}{2} \sum_{i}^3 ({\partial_\mu} {\chi_i})({\partial^\mu} {\chi_i})  = K_1 + K_2 +K_3+...\,,  \, \, \, \, {\rm where}\nonumber\\
& & K_1 = \frac{1}{3 {\cal V}^2} ({\partial_\mu} {\cal V})^2 \sim {\cal O}(1)\,, \nonumber\\
& & K_2 = \frac{3}{2 \sqrt2 {\cal V}} \left\{\frac{\xi_s ({\partial_\mu} \tau_s)^2}{\sqrt{\tau_s}}+\frac{\xi_{sw} ({\partial_\mu} {\tau_{sw}})^2}{\sqrt{\tau_{sw}}}\right\} - \frac{\sqrt2}{{\cal V}} \left(\frac{{\partial_\mu} {\cal V}}{{\cal V}}\right) \biggl\{\xi_s {\sqrt \tau_s} ({\partial_\mu} \tau_s) \nonumber\\
& & \hskip 1cm + \xi_{sw} {\sqrt {\tau_{sw}}} ({\partial_\mu} {\tau_{sw}})- \left(\frac{{\partial_\mu} {\cal V}}{{\cal V}}\right) (\xi_s \, \tau_s^{\frac{3}{2}}+ \xi_{sw} \, {\tau_{sw}}^{\frac{3}{2}})\biggr\} \sim {\cal O}\left(\frac{1}{{\cal V}}\right)\,,\nonumber\\
& &  K_3 = \frac{3 \xi_s \xi_{sw}}{{\cal V}^2 {\sqrt{\tau_s}} {\sqrt{\tau_{sw}}}} \left(\tau_s ({\partial_\mu} {\tau_{sw}}) -\tau_{sw} ({\partial_\mu} {\tau_s})\right)^2 \sim {\cal O}\left(\frac{1}{{\cal V}^2}\right)\,.
\eea
where $\tau_{sw} = \tau_s+\tau_w$. At the leading order, the overall Calabi-Yau volume $\cal V$ is canonically normalized from $K_1$ as
\eq{
{\cal V} \sim \exp\left( {\textstyle {\sqrt{\frac{3}{2}}}}\, \chi_1\right)\,.
}
Utilizing the same, the sub-leading terms appearing as diagonal contributions to $\tau_s$ and ${\tau_{w}}$ result in canonically normalized forms given as
\begin{eqnarray}
& & \hskip -0.7cm \tau_s \sim \frac{1}{\lambda_s} \exp\left(
{\textstyle {\sqrt{\frac{2}{3}}}}\,  \chi_1\right) \, \chi_2^{\frac 4 3} \, , \qquad
{\tau_s+\tau_{w}} \sim \frac{1}{\lambda_{sw}} \exp\left(
{\textstyle {\sqrt{\frac{2}{3}}}}\,  \chi_1\right) \, \chi_3^{\frac 4 3}\,,
\end{eqnarray}
where $\lambda_s = \frac{2^{\frac 7 3} \xi_s^{\frac 2 3}}{3^{\frac 2 3}}$ and
$\lambda_{sw} = \frac{2^{\frac 7 3} \xi_{sw}^{\frac 2 3}}{3^{\frac 2 3}}$ are
constants to be directly read off from the volume form. These aforementioned
canonical forms of ${\cal V}$ and $\tau_s, {\tau_{w}}$ receive corrections at
sub-leading orders. Including ${\cal O}\left(\frac{1}{{\cal V}}\right)$
corrections, we have the following expressions of canonically normalized
(divisor) volume 
moduli
\bea
\label{eq:canonicalmoduli}
& & \hskip -0.6cm {\cal V} \sim \exp\left( {\textstyle
  {{\sqrt{\frac{3}{2}}}}}\,  \chi_1 + {\textstyle \frac{3}{4}} 
   \sum_{i =2}^{3} {\chi_i}^2\right)\,,\nonumber\\
& & \hskip -0.6cm \tau_s \sim \frac{1}{\lambda_s} \exp\left({\textstyle
     {\sqrt{\frac{2}{3}}}}\,  \chi_1\right) \,\left(1+{\textstyle \frac{1}{2}} \sum_{i=2}^{3} {\chi_i}^2\right) \chi_2^{\frac 4 3}\,,\\
& & \hskip -0.6cm \tau_w \sim \exp\left({\textstyle {\sqrt{\frac{2}{3}}}}\,
   \chi_1\right) \, \left(1+{\textstyle \frac{1}{2}} \sum_{i=2}^{3} {\chi_i}^2\right) \left\{\frac{\chi_3^{\frac 4 3}}{{\lambda_{sw}}}- \frac{\chi_2^{\frac 4 3}}{\lambda_s}\right\}\,.\nonumber
\eea
The aforementioned form still receives correction at order ${\cal O}\left(\frac{1}{{\cal V}^2}\right)$ via $K_3$ and in the large volume limit can be safely ignored. 


\section{Eigensystem of the squared-mass matrix}
\label{appendixMass}
Expanding the divisor volume moduli around their respective minima as $\tau_i
= \ov\tau_i + \delta\tau_i$, a generic Lagrangian takes the form,
$${\cal L} = \sum_{ij} K_{ij} (\partial_{\mu}\delta\tau_i) (\partial^{\mu}\delta\tau_j) -
\langle V\rangle -\frac{1}{2} \sum_{ij} V_{ij}\, \delta\tau_i\, \delta\tau_j + {\cal
  O}(\delta\tau^3)\, .$$
Further, the above generic Lagrangian takes a simple form in a basis of canonically normalized moduli which are appropriately normalized via $C^T_a\, K\, C_b=\delta_{ab}$:
$${\cal L} = \frac{1}{2} \sum_{a} (\partial_{\mu}\delta\phi_a)
(\partial^{\mu}\delta\phi_a) - \langle V\rangle -{\frac 1 2}\sum_a M_a^2 \,
\delta\phi_a^2\,.$$ Here $C_{ai}$ and $(M^2)_a$ form the eigensystem of the matrix  ${\cal M}_{ij}=\frac{1}{2} (K^{-1}\, V)_{ij}$. Therefore, it is important to look at the eigensystems of ${\cal M}_{ij}$. The various components of the Hessian matrix are given as under,
\begin{subequations}
\bea
& & \hskip -0.7cm {\bf V}_{\tau_b\tau_b} = \frac{g_s {\xi_b}^{4/3}}{16\pi {\cal V}^{13/3} {\xi_s}} \biggl\{20 \sqrt{2} {\cal V}^2 \sqrt{{\ov\tau_s}} ({a_s} {\ov\lambda_1}-{b_s} {\ov\lambda_2})^2+3 {W_0} {\xi_s} (99 {C_\alpha} {W_0}+128 {\cal V} ({a_s} {\ov\lambda_1}\nonumber\\
& & \hskip 0.5cm -{b_s} {\ov\lambda_2}) {\ov\tau_s})\biggr\} +\frac{g_s \, e^{-a_w \ov\tau_w}}{2\pi {\cal V}^{10/3} {\xi_s}}\biggl\{ \bigl(5 \sqrt{2} {\cal V} ({a_s} {\ov\lambda_1}-{b_s} {\ov\lambda_2})\sqrt{{\ov\tau_s}} (({a_s}-{a_w}) {A_w} {\ov\lambda_1}+({a_w} \nonumber\\
& & \hskip 0.5cm -{b_s}){B_w} {\ov\lambda_2})+48 {W_0} {\xi_s} ({a_s} {A_w} {\ov\lambda_1} {\ov\tau_s}+{a_w} {A_w} {\ov\lambda_1} {\ov\tau_w}-{B_w} {\ov\lambda_2} ({b_s} {\ov\tau_s}+{a_w} {\ov\tau_w}))\bigr) {\xi_b}^{4/3}\biggr\}\nonumber\\
& & \hskip 0.5cm \sim {\cal O}\left(\frac{1}{{\cal V}^{\frac{13}{3}}}\right) + {\cal O}\left(\frac{1}{{\cal V}^{\frac{13}{3}+p}} \right)\,,
\eea
\bea
& & \hskip -0.7cm {\bf V}_{\tau_b\tau_s} = \frac{g_s {\xi_b}^{2/3}}{{4 \sqrt{2} \pi}{\cal V}^{8/3} {\xi_s} \sqrt{{\ov\tau_s}}} \biggl\{{\cal V} ({a_s} {\ov\lambda_1}-{b_s} {\ov\lambda_2}) \bigl(4 {\ov\lambda_1} {\ov\tau_s} {a_s}^2-{\ov\lambda_1} {a_s}+{b_s} {\ov\lambda_2} (1-4 {b_s} {\ov\tau_s})\bigr)\nonumber\\
& & \hskip 0.5cm +{12 \sqrt{2}} {W_0} {\xi_s} \sqrt{{\ov\tau_s}} \bigl({\ov\lambda_1} {\ov\tau_s} {a_s}^2-{\ov\lambda_1} {a_s}+{b_s} {\ov\lambda_2} (1-{b_s} {\ov\tau_s})\bigr)\biggr\}+\frac{g_s \, e^{-a_w \ov\tau_w} {\xi_b}^{2/3}}{2 {\sqrt{2}}\pi{\cal V}^{8/3} {\xi_s} \sqrt{{\ov\tau_s}}} \nonumber\\
& & \hskip 0.5cm \times \biggl\{2 {\cal V} \bigl(2 {a_s}^2 ({a_s}-{a_w}) {A_w} {\ov\lambda_1}^2-({a_s}+{b_s})(({a_s}-{a_w}) {A_w} {b_s}+{a_s} ({b_s}-{a_w}) {B_w}) \nonumber\\
& & \hskip 0.5cm \times {\ov\lambda_2} {\ov\lambda_1}+2 {b_s}^2 ({b_s}-{a_w}) {B_w} {\ov\lambda_2}^2\bigr) {\ov\tau_s}-{\cal V} ({a_s} {\ov\lambda_1}-{b_s} {\ov\lambda_2}) (({a_s}-{a_w}) {A_w} {\ov\lambda_1}+({a_w}\nonumber\\
& & \hskip 0.5cm -{b_s}) {B_w} {\ov\lambda_2}) +6 {\sqrt{2}} {W_0} {\xi_s} \sqrt{{\ov\tau_s}} \biggl({A_w} {\ov\lambda_1} {\ov\tau_s} {a_s}^2+{A_w} {\ov\lambda_1} ({a_w} {\ov\tau_w}-1) {a_s}-{b_s} {B_w} {\ov\lambda_2} \nonumber\\
& & \hskip 0.5cm  \times({b_s} {\ov\tau_s}+{a_w} {\ov\tau_w}-1)\biggr)\biggr\} \nonumber\\
& & \hskip 0.5cm \sim {\cal O}\left(\frac{1}{{\cal V}^{\frac{11}{3}}}\right) + {\cal O}\left(\frac{1}{{\cal V}^{{\frac{11}{3}}+p}} \right)\,,
\eea
\bea
& & \hskip -0.7cm {\bf V}_{\tau_b\tau_w} = \frac{g_s{a_w} \, e^{-a_w \ov\tau_w} {\xi_b}^{2/3}}{\sqrt2 \pi{\cal V}^{8/3} {\xi_s}} \biggl\{{\cal V} ({a_s} {\ov\lambda_1}-{b_s} {\ov\lambda_2}) \sqrt{{\ov\tau_s}}(({a_s}-{a_w}) {A_w} {\ov\lambda_1}+({a_w}-{b_s}) {B_w} {\ov\lambda_2})\nonumber\\
& & \hskip 0.5cm +3 {\sqrt{2}} {W_0} {\xi_s} ({A_w} {\ov\lambda_1} ({a_s} {\ov\tau_s}+{a_w} {\ov\tau_w}-1)-{B_w} {\ov\lambda_2} ({b_s} {\ov\tau_s}+{a_w} {\ov\tau_w}-1))\biggr\}\nonumber\\
& & \hskip 0.5cm \sim {\cal O}\left(\frac{1}{{\cal V}^{\frac{11}{3}+p}} \right)\,,
\eea
\bea
& & \hskip -0.7cm {\bf V}_{\tau_s\tau_s} = \frac{g_s}{24 {\sqrt{2} \pi {\cal V}^2 {\xi_s} {\ov\tau_s}^{3/2}}}\biggl\{6 {2 \sqrt{2}} {W_0} {\xi_s} \bigl({\ov\lambda_1} {\ov\tau_s} {a_s}^3-2 {\ov\lambda_1} {a_s}^2+{b_s}^2 {\ov\lambda_2} (2-{b_s} {\ov\tau_s})\bigr) {\ov\tau_s}^{3/2} \nonumber\\
& & \hskip 0.5cm  +{\cal V} \bigl(-({a_s} {\ov\lambda_1}-{b_s} {\ov\lambda_2})^2-8 \bigl({a_s}^2   {\ov\lambda_1}-{b_s}^2 {\ov\lambda_2}\bigr) {\ov\tau_s} ({a_s} {\ov\lambda_1}-{b_s} {\ov\lambda_2})+8 \bigl(2 {\ov\lambda_1}^2 a_s^4 -{b_s} \nonumber\\
& & \hskip 0.5cm  \times(a_s+b_s)^2 {\ov\lambda_1} {\ov\lambda_2} {a_s}+2 b_s^4 {\ov\lambda_2}^2\bigr) {\ov\tau_s}^2\bigr)\biggr\}+\frac{g_s \, e^{-a_w \ov\tau_w} }{12 {\sqrt{2}} {\cal V}^2 {\xi_s} {\ov\tau_s}^{3/2}} \biggl\{6 {\sqrt{2}} {W_0} {\xi_s} \bigl({A_w}{\ov\lambda_1} {\ov\tau_s}a_s^3 \nonumber\\
& & \hskip 0.5cm +{A_w} {\ov\lambda_1} ({a_w} {\ov\tau_w}-2) {a_s}^2-{b_s}^2 {B_w} {\ov\lambda_2} ({b_s} {\ov\tau_s}+{a_w} {\ov\tau_w}-2)\bigr) {\ov\tau_s}^{3/2}+{\cal V} \bigl(4 \bigl(4 ({a_s}-{a_w}) \nonumber\\
& & \hskip 0.5cm \times {A_w} {\ov\lambda_1}^2 {a_s}^3+4 {b_s}^3 ({b_s}-{a_w}) {B_w} {\ov\lambda_2}^2-({a_s}+{b_s})^2 (({a_s}-{a_w}) {A_w} {b_s}+{a_s} ({b_s}-{a_w}) \nonumber\\
& & \hskip 0.5cm \times {B_w}) {\ov\lambda_1} {\ov\lambda_2}\bigr) {\ov\tau_s}^2-4 \bigl(2 {a_s}^2 ({a_s}-{a_w}) {A_w} {\ov\lambda_1}^2-({a_s}+{b_s}) (({a_s}-{a_w}) {A_w} {b_s}+{a_s} ({b_s}\nonumber\\
& & \hskip 0.5cm -{a_w}) {B_w}) {\ov\lambda_2} {\ov\lambda_1}+2 {b_s}^2 ({b_s}-{a_w}) {B_w} {\ov\lambda_2}^2\bigr){\ov\tau_s}+({b_s} {\ov\lambda_2}-{a_s} {\ov\lambda_1}) (({a_s}-{a_w}) {A_w}  {\ov\lambda_1}\nonumber\\
& & \hskip 0.5cm +({a_w}-{b_s}) {B_w} {\ov\lambda_2})\bigr)\biggr\}\nonumber\\
& & \hskip 0.5cm \sim {\cal O}\left(\frac{1}{{\cal V}^3}\right) + {\cal O}\left(\frac{1}{{\cal V}^{3+p}} \right)\,,
\eea
\bea
& & \hskip -0.7cm {\bf V}_{\tau_s\tau_w} =\frac{g_s {a_w} \, e^{-a_w \ov\tau_w}}{12\pi {\cal V}^2 {\xi_s} \sqrt{{\ov\tau_s}}} \biggl\{\sqrt{2} {\cal V} \biggl(({b_s} {\ov\lambda_2}-{a_s} {\ov\lambda_1}) (({a_s}-{a_w}) {A_w} {\ov\lambda_1}+({a_w}-{b_s}) {B_w}{\ov\lambda_2})\nonumber\\
& & \hskip 0.5cm +2 \bigl(2 {a_s}^2 ({a_s}-{a_w}) {A_w} {\ov\lambda_1}^2-({a_s}+{b_s})
   (({a_s}-{a_w}) {A_w} {b_s}+{a_s} ({b_s}-{a_w}) {B_w}) {\ov\lambda_2} {\ov\lambda_1}\nonumber\\
& & \hskip 0.5cm +2 {b_s}^2 ({b_s}-{a_w}) {B_w} {\ov\lambda_2}^2\bigr) {\ov\tau_s}\biggr)+6 {W_0} {\xi_s} \sqrt{{\ov\tau_s}} \biggl({A_w} {\ov\lambda_1} {\ov\tau_s} {a_s}^2+{A_w} {\ov\lambda_1} ({a_w} {\ov\tau_w}-2) {a_s}\nonumber\\
& & \hskip 0.5cm -{b_s} {B_w} {\ov\lambda_2} ({b_s}{\ov\tau_s}+{a_w} {\ov\tau_w}-2)\biggr)\biggr\}
\sim {\cal O}\left(\frac{1}{{\cal V}^{3+p}} \right)\,,
\eea
\bea
& & \hskip -0.7cm {\bf V}_{\tau_w\tau_w} = \frac{g_s {a_w}^2 \, e^{-a_w \ov\tau_w}}{6\pi {\cal V}^2 {\xi_s}}\biggl\{ \bigl(\sqrt{2} {\cal V} ({a_s} {\ov\lambda_1}-{b_s} {\ov\lambda_2})\sqrt{{\ov\tau_s}} (({a_s}-{a_w}) {A_w} {\ov\lambda_1}+({a_w}-{b_s}) {B_w}{\ov\lambda_2})\nonumber\\
& & \hskip 0.5cm +3 {W_0} {\xi_s} ({A_w} {\ov\lambda_1} ({a_s}
{\ov\tau_s}+{a_w}{\ov\tau_w}-2)-{B_w} {\ov\lambda_2} ({b_s} {\ov\tau_s}+{a_w} {\ov\tau_w}-2))\bigr)\bigr\}\nonumber\\
& & \hskip 0.5cm \sim {\cal O}\left(\frac{1}{{\cal V}^{3+p}} \right)\,.
\eea
\end{subequations}
Utilizing the aforementioned Hessian matrix along with the leading order contribution to the inverse K\"ahler metric (\ref{eq:Ksimp}), the components of the squared-mass matrix defined as ${\cal M}_{ij} = \frac{1}{2} {\cal K}^{-1}_{i \bar k} \,  V_{\bar k j}$ can be easily computed, however, the generic form of respective expressions are too lengthy to present. The volume scalings of the leading terms in the components of ${\cal M}_{ij}$ evaluated at the minimum can be picked up utilizing the large volume limit and the same suffice for our purpose. The squared-mass matrix ${\cal M}_{ij} $ evaluated at the minimum simplifies as below,  

\bea
\label{eq:squaredMassMatrix}
{\cal M}_{ij} = \left(
\begin{array}{lll}
 \frac{f_1}{{\cal V}^{3}}& \hskip 1cm  \frac{f_2}{{\cal V}^{\frac{7}{3}}} & \hskip 1cm \frac{f_3}{{\cal V}^{\frac{7}{3}+p}} \\
 &&\\
 \frac{f_4}{{\cal V}^{\frac{8}{3}}}& \hskip 1cm \frac{f_5}{{\cal V}^{2}} & \hskip 1cm \frac{f_6}{{\cal V}^{2+p}} \\
 &&\\
 \frac{f_7}{{\cal V}^{\frac{8}{3}}}& \hskip 1cm \frac{f_8}{{\cal V}^{2}}& \hskip 1cm \frac{f_9}{{\cal V}^{2+p}}\,
\end{array}
\right)\,,
\eea
where $f_i\in {\mathbb R}$ are order one constants. As expected, in the absence of poly-instanton effects, the upper $2\times2$ block matches with the squared-mass matrix of \cite{Conlon:2007gk} (and subsequently, reproduces the volume scaling estimates for the eigenvalues also) computed in the context of single-hole large volume swiss-cheese model. The aforementioned squared-mass matrix results in hierarchical eigenvalues (in large volume limit) and subsequently, the volume scalings in moduli masses appear to be as under, 
\eq{
M_{\chi_1} \sim {\cal O}(1) \, \frac{M_p}{{\cal V}^{\frac{3}{2}}}\,,\,
M_{\chi_2} \sim  {\cal O}(1) \, \frac{M_p}{{\cal V}}\,, \, M_{\chi_3} \sim
{\cal O}(1) \, \frac{M_p}{{\cal V}^{\frac{2+p}{2}}}\,.
}
For a quick consistency check, one can observe the aforementioned hierarchy in the moduli masses (evaluated at the minimum) just by looking at the trace and determinant of the squared-mass matrix which are as under,
\bea
& & {\rm Tr}[{\cal M}_{ij}] = \sum_{i =1}^3 M_{\chi_i}^2 \sim \frac{{\cal O}(1)}{{\cal V}^2}\,,\quad {\rm and } \quad  {\rm Det}[{\cal M}_{ij}] = \prod_{i =1}^3 M_{\chi_i}^2 \sim \frac{{\cal O}(1)}{{\cal V}^{7+p}}\,
\eea
Furthermore, as the K\"ahler moduli are stabilized in two steps such that $\tau_b,~\tau_s$ are stabilized at order ${\cal V}^{-3}$ and then, the so called Wilson line divisor volume modulus is stabilized via a comparatively sub-dominant contribution and hence, a mass-hierarchy in moduli masses is naturally expected. 


For investigations regarding the computation of the reheating temperature by assuming that the (MS)SM-like model could be supported via wrapping $D7$-branes on the suitable divisor(s), the relevant couplings can be read off from the components of the eigenvectors of the squared-mass matrix. For these estimates, one has to take the numerical approach as it is difficult to read out the volume scalings appearing in the eigenvectors due to (a possible) cancellation among the terms with same volume scalings. We present a numerical calculations for model ${\cal B}_1$ and similar analogous estimates hold for other benchmark models also. The squared-mass matrix evaluated at the minimum is simplified as below,
\bea
\label{eq:squaredMassMatrixNumericalB1}
{\cal M}_{ij} = \left(
\begin{array}{lll}
-2.69 \times 10^{-7} & \hskip 0.5cm 9.42\times 10^{-5}  & \hskip 0.5cm  2.47\times 10^{-8}\\
 &&\\
-4.75\times 10^{-7} & \hskip 0.5cm 1.54\times 10^{-4} & \hskip 0.5cm  3.61\times 10^{-8}\\
 &&\\
4.77\times 10^{-7} & \hskip 0.5cm -1.54\times 10^{-4} & \hskip 0.5cm -3.59\times 10^{-8} \,
\end{array}
\right)\,,
\eea

Comparing \eqref{eq:squaredMassMatrix} and \eqref{eq:squaredMassMatrixNumericalB1}, we find that the coefficients $f_1, f_4, f_8$ and $f_9$ are negative along with $f_6 \simeq |f_9|, \, f_5 \simeq |f_8|, \, f_7 \simeq |f_4| $. These relations which one may not naively expect to be true, hold for all other benchmark models also. The appropriately normalized eigenvectors for estimating divisor volume fluctuations via expressions $\delta\tau_i = \frac{1}{\sqrt2}  \, C_{ai} \, \, \delta\chi_a$ can be written as, 
\bea
\label{eq:Cai}
C_{ai} = \left(
\begin{array}{lll}
-118.73 & \hskip 0.5cm -24.85 & \hskip 0.5cm  -4.82\\
 &&\\
-0.364 & \hskip 0.5cm -40.64 & \hskip 0.5cm  -0.023\\
 &&\\
-5.70 & \hskip 0.5cm 40.74 & \hskip 0.5cm 38.13 \,
\end{array}
\right)\,,
\eea
which is equivalent to the following relations: 
\bea
\label{eq:couplings}
& & \delta{\tau_b} = - 83.955 \, \, \delta\chi_1 -17.568 \, \delta\chi_2  -3.40 \, \, \delta\chi_3\,,\nonumber\\
& & \delta{\tau_s} = -0.258 \, \, \delta\chi_1 - 28.736 \, \,  \delta\chi_2 - 0.0168  \, \, \delta\chi_3\,,\nonumber\\
& & \delta{\tau_w} = -4.031 \, \, \delta\chi_{1} + 28.806 \, \, \delta\chi_2 +  26.961 \, \, \delta\chi_3\,.
\eea

\end{appendix}

\clearpage
\nocite{*}
\bibliography{revPolyCosmo}
\bibliographystyle{utphys}


\end{document}